\documentclass{article}

\usepackage{amsmath,amssymb}
\usepackage{amsthm}
\usepackage{amsbsy}
\usepackage{epsfig,graphics}
\usepackage{booktabs}
\usepackage{comment}
\usepackage{hyperref}
\usepackage{bbm}
\usepackage{cite}
\graphicspath{{./figures/}}

\usepackage{algorithm}
\usepackage{mathtools}
\usepackage{bm}
\usepackage{array,booktabs}
\usepackage{siunitx}

\usepackage{caption} 

\usepackage{cleveref}
\newcommand{\vmax}{\ensuremath{v_\text{max}}}

\def\lse{{l_{\text{se}}}}
\def\kse{{k_{\text{se}}}}

\def\beq{\begin{equation}}
\def\eeq{\end{equation}}

\def\rprime{\hbox{\hskip.25em\raise.5ex\hbox{$'$}\hskip.15em}}

\def\ddn1{{\frac{\partial}{\partial \nu_{\yb}}}}

\newcommand*{\affaddr}[1]{#1} 
\newcommand*{\affmark}[1][*]{\textsuperscript{#1}}

\def\nyu{Courant Institute of Mathematical Sciences, New York University,\\ New York, New York 10012}

\title{A Muscle Crossbridge Theory With Internal Crossbridge Dynamics}

\author{%
Mengjian Hua\affmark[1], Charles S. Peskin\affmark[1]\\
$\,$ \\
\affaddr{\affmark[1]\nyu}\\
}
\date{\today}
\begin{document}

\maketitle

\begin{abstract} 
We describe in this paper a crossbridge model in which an attached
crossbridge behaves like a linear spring with a variable rest length.
We assume in particular that the rest length has a linear
force-velocity relation, and that the force and rest length are both
zero at the moment of crossbridge attachment.  Crossbridges that are
not attached in our model have a fixed probability per unit time of
attachment, and attached crossbridges have a probability per unit time
of detachment that is a function of the crossbridge force.  This
detachment rate is uniquely determined by the requirement that a
limiting form of the model should reproduce the force-velocity curve
and heat of shortening discovered by A.V.Hill~\cite{AVHILL}, and the detachment
rate turns out to be a linearly decreasing function of the crossbridge
force.  The parameters of the model are determined by a fit to steady-
state experimental data; and then an event-driven stochastic simulation
methodology is introduced in order to study the behavior of the
model in a simulated quick-release experiment.  The model explains
how the crossbridge can act like a linear spring on a fast time scale
but have very different properties on a slower time scale.
\end{abstract}

\maketitle
\section{Introduction}

Quantitative understanding of the mechanism of muscle contraction
begins with the work of A.V. Hill~\cite{AVHILL}, who measured and
characterized the velocity of contraction as a function of load, and
who also measured the heat of shortening, so named because the heat
generated (in excess of the maintenance heat) depends only on the
amount of shortening, and not on the rate a which that shortenting
occurs.  The next major contribution was the introduction of the
sliding filament theory, suggested by light microscopy and
confirmed by electron microscopy, that the actin and myosin filaments
of muscle do not change length during contraction, but instead slide
past one another.  What produces this sliding is now understood to be
the crossbridges that protrude from the (thick) myosin filaments and
undergo a cycle in which a crossbridge attaches to a (thin) actin
filament, pulls on that filament in a direction that tends to shorten
the muscle, and then detaches.  Each such crossbridge cycle involves
the hydrolysis of one molecule of ATP, which is needed for the
detachment step.  In the absence of ATP, all of the crossbridges
attach and remain in the attached state, and this leads to the
postmortem phenomenon of rigor.

The first mathematical description of muscle contraction based on the crossbridge cycle was introduced by A.F. Huxley~\cite{huxley1957}. In his theory, the crossbridge is modeled as a linear spring that attaches to a thin filament site in a strained configuration, with the subsequent relaxation of this strain driving muscle contraction.

A landmark series of experiments by Piazzesi~\cite{PIAZZESI2007784} provides a
substantial critique of the Huxley theory.  According to Piazzesi, an
attached crossbridge does indeed act as a linear spring, with a
stiffness constant of  $3.3$ pN/nm, but the force generated by a
crossbridge is fairly constant during the time interval during which
it is attached, even though the crossbridge strain is being relieved
by the relative motion of the thick and thin filaments that occurs
during contraction.  Moreover the energy stored in the spring
immediately after its atachement can be calculated, and it turns
out to be insufficient to account for the work done by the crossbridge
during shortening.

What these observations suggest is that the crossbridge acts as a
linear spring only on a fast time scale, e.g., during quick stretch or
quick release, but that on a slower time scale the crossbridge has
internal dynamics that enable it to maintain a relatively constant
force during the time interval in which it is attached, despite
muscle shortening.

The present paper is an attempt to construct a crossbridge model of
this kind.  We assume that the crossbridge attaches in an unstrained
state, and that strain develops at a finite load-dependent rate, in
competition with the relief of strain by shortening.  In our theory,
the maximum velocity of shortening occurs when the two rates are
equal, so that no strain develops and the force generated by
\textit{each} attached crossbridge is zero.  This is very different
from the Huxley theory, in which the maximum velocity of shortening
occurs when the crossbridge \textit{population} produces zero
\textit{average} force because some crossbridges have been carried
past equilibrium into a state of compression, and these balance the crossbridges that have not yet reached their zero-force configuration and are therefore in a state of tension.

We do not know of other work in which the state of an attached
 crossbridge is characterized by a continuous internal variable with
 its own dynamics, but there are many crossbridge models in which an
 attached (or un-attached) crossbridge can have more than one internal
 state.  An early example is the model of Eisenberg et al. \cite{EISENBERG1980195}, and a
 more recent one is that of Walcott et al.\cite{walcott2012mechanical}.  A model with a
 continuous internal state such as ours can be viewed as a limiting
 case of a model with multiple internal states.

Another way in which our model differs from most is that we treat the
 thin filament as a dense array of binding sites, so that the distance
 to the nearest binding site is not a consideration in our model.  The
 possibility that there might be multiple binding sites within range
 of any un-attached crossbridge was discussed by T.L. Hill \cite{hill1974theoretical}, and
 there is a limiting case of this situation in which the thin filament
 can be regarded as a continuum of binding sites.  This limiting case
 has been used by Lacker and Peskin \cite{lacker1986mathematical} and by Walcott et al.\cite{walcott2012mechanical}, and
 it is also employed in the present paper.  An advantate of this
 formulation is that it allows for the assumption that binding occurs
 in a particular crossbridge configuration, since the configuration in
 which binding occurs does not have to depend on the distance to the
 nearest thin-filament site.  This makes the rate constant for binding
 be a single number, instead of a being a function of distance to the
 nearest site, as in Huxley's original model.  Once it is possible for
 the crossbridge to attach in a particular configuration, it becomes
 possible in our framework to assume that this is a configuration in
 which the crossbridge strain is zero, with strain developing later as
 a result of the internal crossbridge dynamics.

All crossbridge models involve unknown functions, which are typically
 fit to experimental data by trial and error.  Our model has one
 such function, which is the rate contant (probability per unit
 time) for detachment of a crossbridge as a function of the
 crossbridge force.  We determine this function in a systematic
 way, by assuming that a limiting case of our model should account
 for the experimental findings of A.V. Hill \cite{AVHILL} concerning the
 heat of shortening and the force-velocity curve.  This is
 similar to what was done by Lacker \& Peskin \cite{lacker1986mathematical}, who used
 the same experimental findings to identify the unknown functions
 of a different corssbrdge model.  In the present model, we
 find the somewhat paradoxical result that the detachment rate
 is a decreasing function of the force on the crossbridge.

The paper is organized as follows. In Section 2, we introduce our model and
 state its governing equations.  We postpone, however, the discussion of one feature of the model, its series elasticity, until later, since the series elasticity plays no role in our initial considerations.

In Section 3, we derive some steady-state properties of the model in a limiting case, in which the crossbridge stiffness is infinite.  The results of this analysis are in perfect agreement with the force-velocity relation and with the heat of shortening found by A.V. Hill \cite{AVHILL}, provided that we choose a particular rate of crossbridge detachment as a function of the force that is being
 applied to the thin filament by the crossbridge in question, and also provided that we make particular choices for the various parameters of the model.  The rate of detachment as a function of load found in
 this way becomes part of our model definition, and the model parameters found in this way are subsequently used as initial guesses for parameter fitting that does not assume infinite crossbridge stiffness and also does not rely on the results of A.V. Hill~\cite{AVHILL} but
 instead uses the Piazzesi data~\cite{PIAZZESI2007784}.

In Section 4, we prepare for improved parameter fitting by deriving
 exact steady-state results from the governing equations of our model.
 Here, in particular, the crossbridge stiffness is a finite parameter,
 instead of being infinite as in Section 3.  We evaluate the following
 quantities, all as functions of the velocity of shortening: the
 probability that a given crossbridge is attached; the expected value
 of the force generated by each crossbridge; and the expected duration
 of the attached state of a crossbridge, from which with multiplication by the velocity of shortening we get the expected distance traveled by a
 crossbridge during the lifetime of any one of its attached states.
 
 In Section 5, we use the mathematical results of Section 4 to do parameter fitting to the steady-state data of Piazzesi~\cite{PIAZZESI2007784}.

In Section 6, we introduce series elasticity, which is needed for
 simulation of the unsteady behavior of the model, and in particular
 for the simulation of a quick-release experiment.  The methodology
 for the simulation of quick release is explained in Section 7, where
 we introduce a stochastic, event-driven simulation of the active part
 of the crossbridge population within a half sarcomere.  In this
 methodology, individual crossbridges are followed through their
 cycles of attachment, shortening, and detachment, with attachment and
 detachment occurring randomly.  Because the simulation is
 event-driven, there is no time discretization, and the simulation
 generates an exact realization of the stochastic process defined by
 our model.  Results of the simulation of a series of quick-release
 experiments are shown in Section 8, where these simulation results
 are compared to the quick-release experimental results reported by
 Piazzesi~\cite{PIAZZESI2007784}.

In Section 9, we summarize our theory and results together with a discussion of the implications of our model.  

Two appendices are provided: Appendix A explains how the random times
 of crossbridge attachment and detachment are chosen in our simulation
 methodology.  This is standard for attachment, since we assume a
 constant attachment probability per unit time, but it is interesting
 for detachement because the probability per unit time of detachment
 is itself a function of time in our simulations.  Appendix B
 summarizes the results of A.V. Hill~\cite{AVHILL}, and puts them in the form that we use in Section 3.

\section{Muscle Crossbridge Model with Internal Crossbridge Dynamics}
We build a muscle crossbridge model with internal crossbridge dynamics and two time scales. Let an attached crossbridge be characterized by a pair of variables
 $(x,r)$, both of which are functions of the time, $t$.  The variable $x(t)$
 is the distance that the attachment point on the thin filament has
 moved (relative to the thick filament to which the crossbridge
 belongs) from where the attachment point was at the time that
 attachment occurred.  Thus, by definition, $x(t_0)=0$, where $t_0 < t$
 is the time at which attachment occurred.  The variable $r(t)$ is an
 internal variable that plays the role of a rest length, so that
 $r(t)-x(t)$ is the crossbridge strain at time $t$.  We assume that the
 there is no strain at the moment of attachment, and since $x(t_0)=0$
 this implies that $r(t_0)=0$ also. Let the force generated by an attached crossbridge be given by
\begin{equation}
\label{eq:linear_force}
    p(t) = k (r(t) - x(t))
\end{equation}
where $k = 3.3$ pN/nm \cite{PIAZZESI2007784} is the fundamental elastic constant of the myosin motor. Note that $p(t_0)=0$, where $t_0$ is the moment of attachment,
 since we have already assumed that $x(t_0)=r(t_0)=0$. Then, for $t > t_0$ until the moment of detachment, we assume that the crossbridge moves with velocity 
\begin{equation}
\label{eq:length_vel}
    v(t) = \frac{dx}{dt} = - \frac{d l }{dt}
\end{equation}
where $l(t)$ is the length of a half-sarcomere of the muscle at time $t$. For muscle shortening, which is the regime that we are interested in, $\frac{d l}{dt}< 0$ and $v(t)> 0$. 

The internal dynamics of our crossbridge model is defined by
\begin{equation}
\label{eq:r_evolve}
\frac{dr}{dt} = v_\text{max} (1-\frac{p}{p_\infty})
\end{equation}
where $v_\text{max}$ is a parameter that will turn out to be the steady-state velocity with which the half-sarcomere shortens at zero load, and where $p_\infty$ will turn out to be the limiting force
 on an attached crossbridge as $t \rightarrow \infty$ when the muscle
 is isometric, i.e., when $v=0$ \footnote{Note, however, that the
 the crossbridge detaches at some finite time, so the
 approach to $p_\infty$ is cut short by detachment.}. We will determine $p_\infty$ and $v_\text{max}$ later by fitting the model to the experimental data. It follows immediately from equations \eqref{eq:linear_force} - \eqref{eq:r_evolve} that
\begin{equation}
\label{eq:ODE_force}
    \frac{dp}{dt} = k \left(v_\text{max} (1-\frac{p}{p_\infty}) - v\right)
\end{equation}
Assuming that $v$ is independent of time, we can then solve \eqref{eq:ODE_force} directly with initial condition $p(t_0) = 0$, and we obtain
\begin{equation}
\label{eq:force_solution}
    p(t) = p_\infty (1 - \frac{v}{v_\text{max}}) \left(1 - \exp\left({-\frac{k v_\text{max}}{p_\infty} (t - t_0)}\right) \right) 
\end{equation}
and this holds for $t \geq t_0$ until the time of crossbridge detachment. 

Now, we introduce a new parameter $\alpha$ which denotes the probability per unit time of the attachment of an unattached crossbridge. Similarly, we let $\beta (p)$ be the probability per unit time of detachment of an attached crossbridge with force $p$. Note the assumption that $\alpha$ is constant and that $\beta$ is determined by $p$. The value $\alpha$ and the function $\beta(p)$ will be determined later. 

The detachment time $T_\text{d}$ of an attached crossbridge is a random variable
 that is related to $\beta(p)$ in the following way::
\begin{equation}
\label{eq:rate_beta}
    \frac{d \text{Pr} (T_\text{d} > t)}{dt} = - \beta(p(t)) \text{Pr} (T_\text{d} > t)
\end{equation}
where $\text{Pr}$ denotes probability. The initial condition here is $\text{Pr} (T_\text{d} > t_\text{a}) = 1$, where $t_\text{a}$ denotes the time of attachment. 

Similarly, if we let $t_\text{d}$ denote the detachment time of a crossbridge, then its next time of attachment $T_\text{a}$ satisfies
\begin{equation}
\label{eq:rate_alpha}
    \frac{d \text{Pr} (T_\text{a} > t)}{dt} = - \alpha \text{Pr} (T_\text{a} > t)
\end{equation}
with the initial condition $\text{Pr} (T_\text{a} > t_\text{d}) = 1$. The sampling of both $T_\text{d}$ based on equation \eqref{eq:rate_beta} and $T_\text{a}$ based on equation \eqref{eq:rate_alpha} is described in Appendix A. 

\section{Steady-State Results In the Limit $k \rightarrow \infty$}
In the previous section, we formulated the crossbridge model itself but we left the values of the constants $p_\infty, v_\text{max}$, $\alpha$, and the expression for the function $\beta(p)$ to be determined.

The first step towards determining these crossbridge properties
 is to solve the steady state equations of the model in
 a limiting case that turns out to be realistic.  The form of the
 function $\beta(p)$ can then be determined by fitting the limiting
 steady-state results to the fundamental observations of A.V. Hill \cite{AVHILL}
 on the heat of shortening and on the relationship between force and
 velocity for contracting skeletal muscle.  Then, in the following
 section we use more recent experimental data \cite{PIAZZESI2007784} to determine the
 numerical values of the model parameters.

In determining the form of $\beta(p)$, and also to get an initial guess for the subsequent parameter
 fitting, we use a simplified form of the model that is obtained by
 taking the limit $k \rightarrow \infty$.  When all of the model parameters have been
 determined, we will see that the physical value of k is indeed large
 in an appropriate dimensionless sense, and this provides some
 justification for our use of the limit $k \rightarrow \infty$ in the model
 identification process.

An immediate consequence of $k\rightarrow \infty$ is that as soon as a crossbridge attaches, according to equation \eqref{eq:force_solution}, its force jumps immediately from zero to 
\begin{equation}
\label{eq:force_vel_limit}
    p = p_\infty(1-\frac{v}{v_\text{max}})
\end{equation}
Thus $p$ becomes a function of $v$, and the right-hand side of
 equation \eqref{eq:force_vel_limit} will be denoted $p(v)$ in the following.

Let $U(v)$ denote the probability that a crossbridge is attached. Because in this section the crossbridge population is assumed to be
 at a steady state, we have a balance between the rate of attachment of unattached crossbridges and the rate of detachment of attached crossbridges. The balance is as follows
\begin{equation}
\label{eq:balance_rate}
    \alpha (1 - U(v) ) = \beta (p (v) ) U(v) 
\end{equation}
and this is easily solved for $U(v)$ with the result
\begin{equation}
\label{eq:U}
    U(v) = \frac{\alpha}{\alpha + \beta(p(v))}
\end{equation}
Let
\begin{equation}
\label{eq:expected_force}
    P(v) = U(v) p(v)
\end{equation}
which can be interpreted as the expected value of the force exerted by a crossbridge, since the force exerted when a crossbridge is attached is $p(v)$, and
 when a crossbridge is not attached its force is zero.

We can also evaluate the mean cycling time of a crossbridge at the steady state, and we denote it by $\overline{T_\text{c}} (v)$, which is given by
\begin{equation}
\label{eq:mean_cycling_0}
    \overline{T_\text{c}} (v) = \frac{1}{\alpha} + \frac{1}{\beta(p(v))} 
\end{equation}
Note that  $1/\overline{T_\text{c}}(v)$ is the rate of crossbridge cycling, per crossbridge,
 i.e., the mean number of cycles per unit time that any particular
 crossbridge undergoes.  This is given by either side of equation \eqref{eq:balance_rate}, so we have the equation 
\begin{equation}
\label{eq:mean_cycling_time}
    \frac{1}{\overline{T_\text{c}} (v)} = U(v) \beta(p(v))
\end{equation}
From A.V.Hill \cite{AVHILL}, we have the following two equations:
\begin{equation}
\label{eq:Hill1}
P(v) = P(0) \frac{1 - \frac{v}{v_\text{max}}}{1 + \frac{4v}{v_\text{max}}} 
\end{equation}
\begin{equation}
\label{eq:Hill2}
    \frac{1}{\overline{T_\text{c}} (v)} = \frac{1}{\overline{T_\text{c}} (0)}\frac{1 + \frac{20v}{v_\text{max}}}{1 + \frac{4v}{v_\text{max}}} 
\end{equation}
For details on how these equations summarize Hill’s results, see Appendix B.  Note that we are not bothering to distinguish between
 the parameter $\vmax$ of our model and the parameter $\vmax$ of the Hill
 force-velocity curve, since it is obvious that these two parameters
 have to agree if $P(v)$ as given by \eqref{eq:Hill1} is to be zero at the same
 value of v as P(v) as given by \eqref{eq:expected_force}, see also equation \eqref{eq:force_vel_limit}.

Now we divide \eqref{eq:Hill2} by \eqref{eq:Hill1}, divide \eqref{eq:mean_cycling_time} by \eqref{eq:expected_force},
 and set the two resulting right-hand sides equal to each other
 since the left-hand sides in both cases are equal to $1/(\overline{T_\text{c}}(v)P(v))$:
\begin{equation}
    \frac{1}{\overline{T_\text{c}} (0) P(0) } \frac{1 + \frac{20v}{v_\text{max}}}{1 - \frac{v}{v_\text{max}}} = \frac{1}{\overline{T_\text{c}} (v) P(v) } = \frac{\beta (p(v))}{p (v)}
\end{equation}
Therefore, we have 
\begin{equation}
\label{eq:intermediate_beta}
    \beta(p(v)) = p(v) \frac{1}{\overline{T_\text{c}} (0) P(0) } \frac{1 + \frac{20v}{v_\text{max}}}{1 - \frac{v}{v_\text{max}}}
\end{equation}
Substituting \eqref{eq:force_vel_limit} into \eqref{eq:intermediate_beta} yields
\begin{equation}
\label{eq:beta_plug}
    \beta(p(v)) = \frac{p_\infty}{\overline{T_\text{c}} (0) P(0)} \big(1 + 20(1 - \frac{p(v)}{p_\infty})\big)
\end{equation}
From this result and \eqref{eq:force_vel_limit}, it follows that $p(0) = p_\infty$. Therefore, setting v=0
 in equation \eqref{eq:beta_plug} we get
\begin{equation}
\label{eq:beta_p_infty}
    \beta(p_\infty) = \frac{p_\infty}{\overline{T_\text{c}} (0) P(0)}
\end{equation}
and this shows that equation \eqref{eq:beta_plug} can also be written as
\begin{equation}
\label{eq:analytical_beta}
    \beta(p(v)) = \beta(p_\infty) \big(1 + 20(1 - \frac{p(v)}{p_\infty})\big)
\end{equation}
Combining this result with \eqref{eq:U}, \eqref{eq:expected_force}, and \eqref{eq:mean_cycling_0}, it follows that
\begin{equation}
\label{eq:U_k_infty}
    U(v) = \frac{\alpha}{\alpha + \beta(p_\infty) \big(1 + 20(1 - \frac{p(v)}{p_\infty})\big)} = \frac{\alpha}{\alpha + \beta(p_\infty) \big(1 + \frac{20 v}{\vmax} \big)} 
\end{equation}
\begin{equation}
\label{eq:force_3}
     P(v)= \frac{\alpha p_\infty (1- \frac{v}{v_\text{max}})}{\alpha + \beta(p_\infty) \big(1 + 20(1 - \frac{p(v)}{p_\infty})\big)} = \frac{\alpha p_\infty (1- \frac{v}{v_\text{max}})}{\alpha + \beta(p_\infty) \big(1 + \frac{20 v}{\vmax})\big)}
\end{equation}
\begin{equation}
\label{eq:T_c_3}
    \frac{1}{\overline{T_\text{c}}(v)} = \frac{\alpha \beta(p_\infty) \big(1 + 20(1 - \frac{p(v)}{p_\infty})\big)}{\alpha + \beta(p_\infty) \big(1 + 20(1 - \frac{p(v)}{p_\infty})\big)} = \frac{\alpha \beta(p_\infty) \big(1 + \frac{20 v}{\vmax}\big)}{\alpha + \beta(p_\infty) \big(1 + \frac{20 v}{\vmax}\big)}
\end{equation}
in which we have made use of equation (8)
to replace $(1 - \frac{p(v)}{p_\infty})$ by $\frac{v}{\vmax}$.

Now we require \eqref{eq:T_c_3} to agree with \eqref{eq:Hill2}, and \eqref{eq:force_3} to agree with \eqref{eq:Hill1}.In each case, there is a common factor that cancels, and we are left with the two equations:

\begin{equation}
\label{eq:solve_1_alpha}
    \frac{\alpha \beta(p_\infty) }{\alpha + \beta(p_\infty) \big(1 + \frac{20 v}{\vmax}\big) } = \frac{1}{\overline{T_\text{c}} (0)}\frac{1}{1 + \frac{4v}{v_\text{max}}} 
\end{equation}

\begin{equation}
\label{eq:solve_2_alpha}
    \frac{\alpha p_\infty }{\alpha + \beta(p_\infty) \big(1 + \frac{20 v}{\vmax})} = P(0) \frac{1}{1 + \frac{4v}{v_\text{max}}} 
\end{equation}

Equation \eqref{eq:solve_1_alpha} and equation \eqref{eq:solve_2_alpha} are supposed to hold for all $v$. Making use of this and also equation \eqref{eq:beta_p_infty} we get the simple results
\begin{equation}
\label{eq:beta_p_infty_alpha}
    \beta(p_\infty) = \frac{1}{4} \alpha
\end{equation}
\begin{equation}
    \frac{1}{\overline{T_\text{c}} (0) } = \frac{1}{5} \alpha
\end{equation}
and it follows immediately from \eqref{eq:analytical_beta} and \eqref{eq:beta_p_infty_alpha} that 

\begin{equation}
\label{eq:beta_final}
    \beta(p) = \frac{\alpha}{4} \big(1 + 20(1 - \frac{p}{p_\infty})\big)
\end{equation}
With $\beta(p_\infty)$ given by \eqref{eq:beta_p_infty_alpha}, we can evaluate the right-hand sides of equations \eqref{eq:U_k_infty}-\eqref{eq:T_c_3}
and thereby obtain the following results:
\begin{equation}
\label{eq:U(v)_infinite}
    U(v) = \frac{ 4 \vmax}{5 \vmax + 20 v}
\end{equation}
\begin{equation}
\label{eq:force_fit_eq}
    P(v) = \frac{4 p_\infty}{5}\frac{1 - \frac{v}{\vmax}}{1 + 4\frac{v}{\vmax}}
\end{equation}
\begin{equation}
\label{eq:mean_cycle}
	\overline{T_\text{c}} (v) = \frac{ \alpha \vmax}{5 \vmax + 20 v} \big(1 + \frac{20 v}{\vmax}\big)
\end{equation}

Moreover, we let $\overline{T_\text{a}} (v) $ be the mean duration of the attached state of a crossbridge. Since the expected mean duration of the unattached state is $\frac{1}{\alpha}$. We know that 
\begin{equation}
\label{eq:mean_duration_U}
    U(v) = \frac{\overline{T_\text{a}} (v)}{\overline{T_\text{a}} (v) + \frac{1}{\alpha}}
\end{equation}
because $U(v)$ by definition is the probability that any particular crossbridge is attached. By combining \eqref{eq:mean_duration_U} with \eqref{eq:U(v)_infinite} we therefore get the result that
\begin{equation}
\label{eq:fit_mean_duration_0}
    \overline{T_\text{a}}(v) = \frac{4 }{\alpha (1 + 20\frac{v}{\vmax})}
\end{equation}
 
This result can also be obtained by subtracting $1/\alpha$ from
 the mean cycle time as given by \eqref{eq:mean_cycle}. It follows that the mean step length of an attached crossbridge is 
 \begin{equation}
 	\label{eq:fit_mean_step_length}
 	\overline{S_\text{a}}(v) = v \overline{T_\text{a}}(v) = \frac{4v }{\alpha (1 + 20\frac{v}{\vmax})}
 \end{equation}
 
 The results derived in this section will be used in the following
 ways.  First, equation \eqref{eq:beta_final} will be used from now on as our formula
 for $\beta(p)$.  Also, when it comes to parameter fitting, we will
 use the results of this section to obtain a good initial guess that
 will then be improved by making use of the exact results obtained
 in the next section.
 
 Our use of equation \eqref{eq:beta_final} for $\beta(p)$ requires discussion, since \eqref{eq:beta_final} was derived by considering the limit $k \rightarrow \infty$, but we will be
 using the same formula for $\beta(p)$ together with the finite, measured
 value of $k$.  This somewhat questionable way of proceeding can perhaps
 be justified in the following way.
 
 Note that there are $\textit{two}$ approximations made in the above
 derivation of equation \eqref{eq:beta_final} for $\beta(p)$, one on the modeling side
 i.e., the use of the limit $k \rightarrow \infty$, and the other on the
 experimental side, i.e. the use of A.V.Hill's 1938 description of his
 experimental results.  This description is only approximate, and was later
revised by Hill himself with regard to the heat
of shortening~\cite{hill1964effect}, and by others with regard to
the force-velocity curve (reviewed in~\cite{alcazar2019shape}).  The revised descriptions are
 considerably more complicated and less beautiful.  It seems plausible
 to us that Hill's 1938 summary of his results  may be simple (and therefore
 beautiful) precisely because it corresponds $\textit{exactly}$ to the
 behavior of muscle in some limiting case, like the limit $k \rightarrow \infty$ that we have chosen to consider.  Indeed, we have shown in this
 section that the present model, with equation \eqref{eq:beta_final} for $\beta(p)$ and in
 the limit $k \rightarrow \infty$, has precisely the force-velocity curve and
 heat of shortening that Hill chose to use in his 1938 summary of his
 experimental observations \cite{AVHILL}.
 
 A somewhat different way of thinking about equation \eqref{eq:beta_final} it is that it
 is simply an additional hypothesis that is part of the definition of
 our model.  From this point of view, the fact that we recover Hill's
 1938 results when we make this hypothesis together with the limit $k \rightarrow \infty$ is irrelevant, since we will later be testing the full model,
 including both equation \eqref{eq:beta_final} for $\beta(p)$ and also the finite, measured
 value of $k$, against a more recent and more comprehensive data set.
 
 \section{Exact Steady-State Results for $k < \infty$}
 In this section we derive exact formulae for the steady state of our
 model with k finite.  By steady state we mean that the each half-sarcomere
 is shortening at a constant velocity v, and that the crossbridge population
 has had enough time to equilibrate to this velocity of shortening.

Now, with $k$ being finite, the muscle crossbridge force $p$ is not a function only of $v$, and it is described as in \eqref{eq:force_solution},  Note, however, that in equation \eqref{eq:force_solution}, $t_0$ is the time of attachment
 of a crosssbridge, and this in general will be different for each
 crossbridge that happens to be attached at the time $t$.  Thus, we need
 to consider the crossbridge population as a whole instead of each
 individual crossbridge.  Accordingly, we let $u(v,p)$ be a probability density function such that $\int_{p_1}^{p_2} u(v,p) dp$ denotes the probability that any particular crossbridge is attached and has $p \in (p_1,p_2)$. Then, we define
\begin{equation}
\label{eq:U(v)_finite}
     U(v) = \int_{0}^{p_\infty(1- \frac{v}{v_\text{max}})} u(v,p) dp
 \end{equation}
 and $U(v)$ is the probability that any particular crossbridge is attached. Also, we define
 \begin{equation}
 \label{eq:P(v)_finite}
     P(v) = \int_{0}^{p_\infty(1- \frac{v}{v_\text{max}})} p u(v,p) dp
 \end{equation}
 as the expected value of the force generated by any particular crossbridge. Note that this averaging is over all possible configurations of a crossbridge, including the case in which it is unattached with zero force. The expected force conditioned on attachment is thus 
 $P(v)/U(v)$. 

With the definition of probability density $u(v,p)$, we can rewrite the balance equation \eqref{eq:balance_rate} as follows:
 \begin{equation}
 \label{eq:balance_rate_2}
     \alpha (1 - U(v) ) = \int_{0}^{p_\infty(1- \frac{v}{v_\text{max}})} \beta(p) u(v,p) dp
 \end{equation}
 Since we know what $\beta(p)$ is from equation \eqref{eq:beta_final}, we conclude that
 \begin{equation}
 \label{eq:plug_in_beta}
     \alpha (1 - U(v) ) = \int_{0}^{p_\infty(1- \frac{v}{v_\text{max}})} \frac{\alpha}{4} \big(1 + 20(1 - \frac{p}{p_\infty})\big) u(v,p) dp
 \end{equation}
 The parameter $\alpha$ cancels, and then the integral on the right-hand
 side of \eqref{eq:plug_in_beta} is easily expressed in terms of U(v) and P(v), which
 are defined by equations \eqref{eq:U(v)_finite}-\eqref{eq:P(v)_finite}.  Solving for $P(v)$, we then find
\begin{equation}
 \label{eq:fit_force_num}
     P(v) = p_\infty (\frac{5}{4} U(v) - \frac{1}{5})
\end{equation}
Note that this is a very specific and simple prediction of a
 relationship between the expected crossbridge force $P(v)$ ahd the
 probability $U(v)$ that a crossbridge is attached.

Moreover, from \eqref{eq:balance_rate_2}, we have the integral equation
 \begin{equation}
 \label{eq:integral_eq}
     \alpha (1 - U(v) ) = \int_{0}^{p} \beta(p') u(v,p') dp' + u(p,v) \frac{dp}{dt} (v,p)
 \end{equation}
 In this equation $\frac{dp}{dt}(v,p)$ is the rate of change of force of
 an individual crossbridge with force p when the shortening
 velocity of the half-sarcomere is $v$.  This is given by
 equation \eqref{eq:ODE_force} as follows:
 \begin{equation}
     \frac{dp}{dt} (v,p) = k \left(v_\text{max} (1-\frac{p}{p_\infty}) - v\right)
 \end{equation}
 
Differentiating \eqref{eq:integral_eq} with respect to $p$ gives
 \begin{equation}
 \label{eq:diff_q_u}
     0 = \beta(p) u(v,p) + k\frac{\partial u}{\partial p} (p,v) \left(v_\text{max} (1-\frac{p}{p_\infty}) - v\right) - \frac{k \vmax u(p,v)}{p_\infty} 
 \end{equation}
With $\beta(p)$ given by \eqref{eq:beta_final}, this becomes the following ODE for u(v,p):
 \begin{equation}
 \label{eq:ode_u_finite}
      \frac{p_\infty}{u(v,p)} \frac{\partial u}{\partial p} (v,p) =  
      \frac{1 - \frac{\alpha p_\infty}{4 k \vmax} (1 + 20 \frac{v}{\vmax})}{1 - \frac{p}{p_\infty} - \frac{v}{\vmax}} - \frac{5 \alpha p_\infty}{k \vmax}
  \end{equation}
  Integrating \eqref{eq:ode_u_finite} gives 
  \begin{equation}
  	u(v,p) = u(v,0) \exp(-5 \frac{\alpha p_\infty
  	}{k \vmax} \frac{p}{p_\infty} )(1 - \frac{p}{p_\infty (1 - \frac{v}{\vmax})})^{(\frac{\alpha p_\infty
  	}{k \vmax} (1+ 20\frac{v}{\vmax}) - 1 )}
  \end{equation}

To determine $u(v,0)$, we can set $p = 0$ in equation \eqref{eq:integral_eq}, which gives
 \begin{equation}
     \alpha (1- U(v) ) = k u(v,0) \left(v_\text{max} - v\right)
 \end{equation}
 and thus 
 \begin{equation}
     u(v,0) = \frac{\alpha (1- U(v) )}{\vmax - v}
 \end{equation}
 
 To summarize, we obtain the following expression for $u(v,p)$: 
 
 \begin{equation}
 	u(v,p) = \frac{\alpha (1- U(v) )}{\vmax - v} \exp(-5 \frac{\alpha p_\infty
  	}{k \vmax} \frac{p}{p_\infty} )(1 - \frac{p}{p_\infty (1 - \frac{v}{\vmax})})^{(\frac{\alpha p_\infty
  	}{k \vmax} (1+ 20\frac{v}{\vmax}) - 1 )}
 \end{equation}
 
 To clean up the notation, we let 
 \begin{equation}
 \label{eq:epsilon}
 	\epsilon = \frac{\alpha p_\infty}{k \vmax}
 \end{equation}
 and make the change of variables 
 \begin{equation}
 	q = \frac{p}{p_\infty (1- \frac{v}{\vmax})}
 \end{equation}
 \begin{equation}
 	dq = \frac{d p}{p_\infty (1- \frac{v}{\vmax})}
 \end{equation}
 
 The domain of integration with respect to $q$ is then $[0,1]$ and we can rewrite \eqref{eq:integral_eq} as 
 \begin{equation}
 \label{eq:U_I}
 	U_\epsilon (v) = (1 - U_\epsilon (v) ) I_\epsilon (v)
 \end{equation}
where we are now making explicit the dependence of $U(v)$ on $\epsilon$ and where 
\begin{equation}
\label{eq:I_epsilon}
	I_\epsilon (v) = \epsilon \int_{0}^{1} \exp(-5 \epsilon (1-\frac{v}{\vmax})q ) (1-q)^{\frac{\epsilon}{4} (1 + 20\frac{v}{\vmax}) - 1} dq
\end{equation}

The integrand in equation \eqref{eq:I_epsilon} typically blows up at $q = 1$, since
 the power of $1-q$ is typically negative.  This singularity is
 integrable, however, since the power of $1-q$ is always greater
 than $-1$.  Indeed, we can remove the singularity by using integration
 by parts to rewrite $I_\epsilon(v)$ as follows:
 \begin{equation}
  \label{eq:J_in_I}
	I_\epsilon (v) = \frac{4}{1 + 20 \frac{v}{\vmax} } (1 - 5 \epsilon (1 - \frac{v}{\vmax} ) J _\epsilon (v) )
\end{equation}
 where 
 \begin{equation}
 \label{eq:J}
	J_\epsilon (v) = \int_0^1 \exp(-5 \epsilon (1- \frac{v}{\vmax})q ) (1-q)^{\frac{\epsilon}{4} (1 + 20 \frac{v}{\vmax})} dq 
\end{equation}

We then have the following steady state results: 
\begin{itemize}
	\item From \eqref{eq:U_I} and \eqref{eq:J_in_I}
	\begin{equation}
		U_\epsilon (v) = \frac{I_\epsilon(v)}{1+ I_\epsilon(v)} = \frac{4(1 - 5 \epsilon (1 - \frac{v}{\vmax} ) J _\epsilon (v))}{5( 1+ 4\frac{v}{\vmax} - 4 \epsilon (1 - \frac{v}{\vmax} ) J _\epsilon (v))}
	\end{equation}
	\item 
	
	Then, from \eqref{eq:fit_force_num},  
	\begin{equation}
		P_\epsilon(v) = p_\infty (\frac{5}{4} U(v) - \frac{1}{5}) = \frac{(4 - 21\epsilon J _\epsilon (v)) (1- \frac{v}{\vmax}) }{5( 1+ 4\frac{v}{\vmax} - 4 \epsilon (1 - \frac{v}{\vmax} ) J _\epsilon (v))}
	\end{equation}
	
	\item Also, we know from \eqref{eq:mean_duration_U} that mean duration of the attached state of a crossbridge can be expressed as
	\begin{equation}
		\overline{T_\text{a}}(v) = \frac{U_\epsilon (v)}{\alpha (1-U_\epsilon (v))} = \frac{4(1 - 5 \epsilon (1 - \frac{v}{\vmax} ) J _\epsilon (v))}{\alpha (1 + 4 \frac{v}{\vmax})}
	\end{equation}
	\item It follows that the mean step length of a crossbridge is
	\begin{equation}
	\label{eq:mean_step_length}
		(\overline{S_a})_\epsilon (v) = v \overline{T_\text{a}}(v) = \frac{4v (1 - 5 \epsilon (1 - \frac{v}{\vmax} ) J _\epsilon (v))}{\alpha (1 + 4 \frac{v}{\vmax})}
	\end{equation}
	\end{itemize}

We claim that the asymptotic behavior as $k \rightarrow \infty$ (i.e. $\epsilon \rightarrow 0$) of the model presented in this section is exactly the same as the steady-state behavior derived by assuming
 that limiting case from the beginning, as was done in Section 3.

To prove this,  Note that $J_\epsilon(v)$ is bounded from above by
\begin{equation}
	J_\epsilon(v) \leq \int_0^1 (1-q)^{\frac{\epsilon}{4} (1 + 20 \frac{v}{\vmax})} dq = \frac{1}{1 + \frac{\epsilon}{4} (1 + 20 \frac{v}{\vmax})}
\end{equation}
and bounded from below by 
\begin{equation}
	J_\epsilon(v) \geq (1 -5 \epsilon (1- \frac{v}{\vmax}))\int_0^1 (1-q)^{\frac{\epsilon}{4} (1 + 20 \frac{v}{\vmax})} dq = \frac{1 -5 \epsilon (1- \frac{v}{\vmax})q }{1 + \frac{\epsilon}{4} (1 + 20 \frac{v}{\vmax})}
\end{equation}
It follows from these bounds that 
\begin{equation}
	J_\epsilon(v) = 1 + \text{O}(\epsilon)
\end{equation}
Therefore, we have the following estimate for $I_\epsilon (v)$: 
\begin{equation}
\label{eq:asym_I}
	I_\epsilon (v) = \frac{1}{1 + 20 \frac{v}{\vmax} } \left(1 - 5 \epsilon (1 - \frac{v}{\vmax} ) + \text{O}(\epsilon^2)\right)
\end{equation}
From \eqref{eq:asym_I} and \eqref{eq:U_I},
\begin{equation}
\label{eq:U_I_asym}
	U_\epsilon (v) = \frac{I_\epsilon (v)}{1 + I_\epsilon (v)} = \frac{4}{5} \frac{1 - \epsilon \frac{(1 - \frac{v}{\vmax}) (1 + 20\frac{v}{\vmax})}{1 + 4\frac{v}{\vmax}} + \text{O} (\epsilon^2) }{1 + 4\frac{v}{\vmax}}
\end{equation}
Then, from \eqref{eq:U_I_asym} and \eqref{eq:fit_force_num}, 
\begin{equation}
\label{eq:P_I_asym}
	P_\epsilon (v) = \frac{4}{5} p_\infty (\frac{1 - \frac{v}{\vmax}}{1 + 4 \frac{v}{\vmax}}) (1 - \frac{5\epsilon}{4} \frac{1 + 20 \frac{v}{\vmax}}{1 + 4 \frac{v}{\vmax}} + \text{O} (\epsilon^2))
\end{equation}
Moreover, from \eqref{eq:mean_step_length},
\begin{equation}
\label{eq:mean_step_length_app}
	(\overline{S_a})_\epsilon (v) = \frac{4v (1 - 5 \epsilon (1 - \frac{v}{\vmax} )) + \text{O} (\epsilon^2)}{\alpha (1 + 4 \frac{v}{\vmax})}
\end{equation}
If we let $\epsilon \rightarrow 0$ (i.e. $k \rightarrow \infty$) in \eqref{eq:U_I_asym}-\eqref{eq:mean_step_length_app}, these three equations will reduce to \eqref{eq:U(v)_infinite},\eqref{eq:force_fit_eq} and \eqref{eq:fit_mean_step_length} and this justifies the procedure used in Section 3 to obtain these limiting results.

\section{Parameter Fitting}

The steady-state analysis presented in Section 3 gives us equation \eqref{eq:beta_final} for $\beta (p)$. In this section, we determine the parameter values for the constants $p_\infty, \vmax$, and $\alpha$. For this purpose we use data from \cite{PIAZZESI2007784}.  Note, however, that in
 extracting data from this paper we avoid the data points
 corresponding to the isometric case, in which the muscle is not
 actually shortening.  There seems to be something fundamentally
 different about this case that is not captured by our model (or by
 any model of which we are aware, see \cite{alcazar2019shape} for a discussion of this
 issue).  This shows up most clearly in Figures 3D and 4B of \cite{PIAZZESI2007784}, in
 which the force per attached crossbridge is plotted as a function of
 filament load (in Figure 3D), and as a function of velocity (in
 Figure 4B).  In both cases the data seem to fall on a straight line
 except for the isometric data point which deviates very markedly from
 that line.  In these data, the apparent limit as $v \rightarrow 0$ of the
 force per attached crossbridge is $1.25$ times the isometric force per
 attached crossbridge.  To avoid this issue, we fit our model only to
 data in which the muscle is actually shortening.

 Another issue that arises in the use of the data from \cite{PIAZZESI2007784} is that
 Figure 4A of \cite{PIAZZESI2007784} is a plot of the \textit{number} of attached
 crossbridges as a function of the velocity of shortening.  Our model,
 however, makes predictions about the \textit{probability} of
 crossbridge attachment, denoted $U(v)$, which is macroscopically
 equivalent to the \textit{fraction} of attached crossbridges.  To
 make a connection between the data and our theory, we introduce a
 parameter $N_c$, which is the number of myosin crossbridges in a
 half-sarcomere that are participating in the crossbridge cycle.  This
 number might be less than the number of myosin crossbridges in a
 half-sarcomere (which is stated in \cite{PIAZZESI2007784} to be 294), because only some
 of the myosin crossbridges may be adjacent to activated thin
 filament.  In this interpretation, the parameter $N_c$ should be
 dependent on the state of activation of the muscle, i.e., on the
 concentration of intracellular free Ca$^2+$ in the neighborhood of
 the contractile machinery.  We assume, however, that $N_c$ is
 constant for a muscle in a given state of activation, and in
 particular that it was constant throughout the experiments reported
 in \cite{PIAZZESI2007784}.  The particular value of $N_c$ will then be determined as
 part of the parameter-fitting process described below.  What we find in this way is that, under the conditions of the
 experiments described in~\cite{PIAZZESI2007784}, the number of cycling myosin
 crossbridges is about 3/8 of the total number of myosin crossbridges
 that are present in the muscle.. Our parameter-fitting procedure is as follows:

\begin{itemize}
    
 \item \textbf{Getting the initial guess for the parameters}: A key observation we have made during the parameter fitting
 is that the fit is not sensitive to $\epsilon$. In other words, we find that any $\epsilon \in [0,0.15]$ will give an almost equally good fit. Moreover, the fitting of $\epsilon$ is sensitive to the initial guess of the parameters. So, in order to resolve this inexactness in the choice of $\epsilon$, we let the initial guess be the ones we would obtain from the $\epsilon \rightarrow 0$ limit, which is the most reasonable initial guess we can make. 
	From Section 4, we have the expressions for $P(v)$ in \eqref{eq:force_fit_eq} and $\overline{S_\text{a}}(v)$ in \eqref{eq:fit_mean_step_length}. Moreover, we let $N_\text{a}(v) = N_c U(v)$, where $U(v)$ is defined by \eqref{eq:U(v)_infinite}. Then, we can use these expressions to fit the parameters from data in Figure 1C, 4A, 4C of \cite{PIAZZESI2007784} and get estimates of $N_c,\vmax,p_\infty,\alpha$. Even though we start by assuming $\epsilon = 0$, we can still get a value of $\epsilon$ from \eqref{eq:epsilon}, since the value of $k$ is known
 to be $3.3 pN/nm$. We use the mean-square error as the objective function for optimizing the parameter values and we use the Levenberg-Marquardt method for performing this nonlinear least-squares optimization.
 
The parameters obtained in this way are listed in the right-hand side of the arrows in first row of Table ~\ref{table:parameters}, and these will be used as the initial guess for the next
 step.
 
 \item \textbf{Fitting $p_\infty, \epsilon, N_c, \vmax$}: now we replace $U(v)$ from \eqref{eq:U(v)_infinite} by $U_\epsilon(v)$ in \eqref{eq:U_I} and keep the relation $N_\text{a}(v) = N_c U_\epsilon(v)$. We use it together with \eqref{eq:fit_force_num} to fit the parameters with the data in Figure 1c and Figure 4a of \cite{PIAZZESI2007784} with the same methodology as in the previous step. This gives the values of $p_\infty, \epsilon, N_c, \vmax$, which are all we need except for $\alpha$.
 \item \textbf{Fitting $\alpha$}:
Because of the uncertainty in $\epsilon$, we choose not to compute $\alpha$ with \eqref{eq:epsilon} through $\alpha = \frac{k \vmax}{p_\infty \epsilon}$. Instead, we use the values of $p_\infty, \vmax$ from the previous step and use \eqref{eq:epsilon} to eliminate $\epsilon$ in \eqref{eq:mean_step_length}. Then, the only parameter remains to be determined in  \eqref{eq:mean_step_length} is $\alpha$. We fit the data in Figure 4C of \cite{PIAZZESI2007784} to \eqref{eq:mean_step_length} and we obtain our final estimate for $\alpha$, which is also given in Table~\ref{table:parameters}. \end{itemize}

\begin{table}[ht]
\centering
\begin{tabular}{ccccc}
\hline
{Parameter values}  & $\alpha$ ($s^{-1}$) & $p_\infty$ (pN) & $\vmax$ (nm/s) & $N_c$ \\ \hline
$k = \infty$   &       $60 \rightarrow 65.4$        &  $6.0 \rightarrow 9.24$ & $2\times 10^3 \rightarrow 2.24 \times 10^3$ & $100 \rightarrow 131$                 \\
$k = 3.3$ pN/nm \cite{PIAZZESI2007784}    &     $68.2$                  & $9.98$   &       $2.75\times 10^3$    & $116$           \\                  \hline
\end{tabular}
\caption{The parameter values fitted with the steady-state theory of $k\rightarrow \infty$ and the exact steady-state theory with $k = 3.3$ pN/nm. On the first row, the values on the left of the arrows are the initial guesses and the values on the right of the arrows are the fitted paramters with the steady-state theory with $k = \infty$. The second row is obtained by performing paramter fitting with $k = 3.3$ pN/nm with the values obtained in the first row being the warm-starts for the optimization. Note that the values on the second row of the table give $\epsilon = 0.08$,
 which is indeed small.  This justifies our claim that the stiffness parameter k is large in an appropriate dimensionless sense.  }
\label{table:parameters}
\end{table}

\begin{figure}[t]
  \centering
   \includegraphics[scale=0.19]{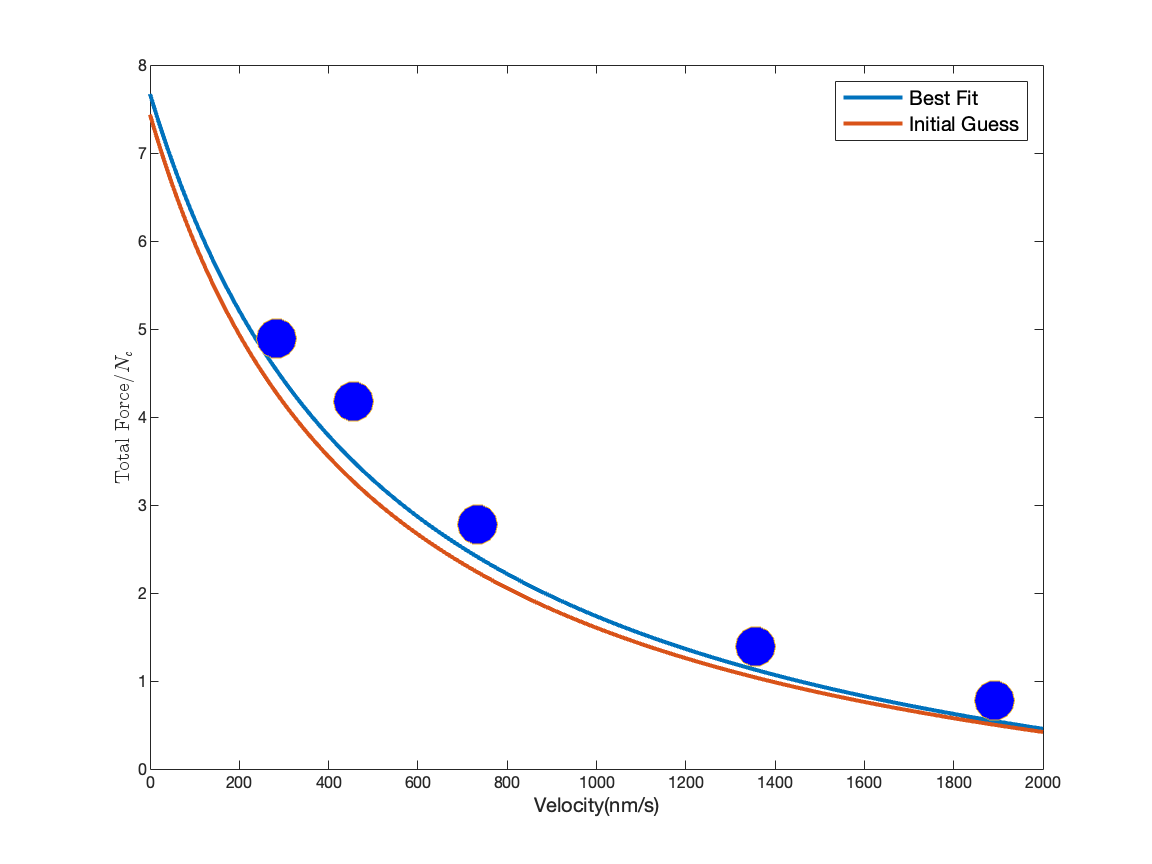}
   \includegraphics[scale=0.19]{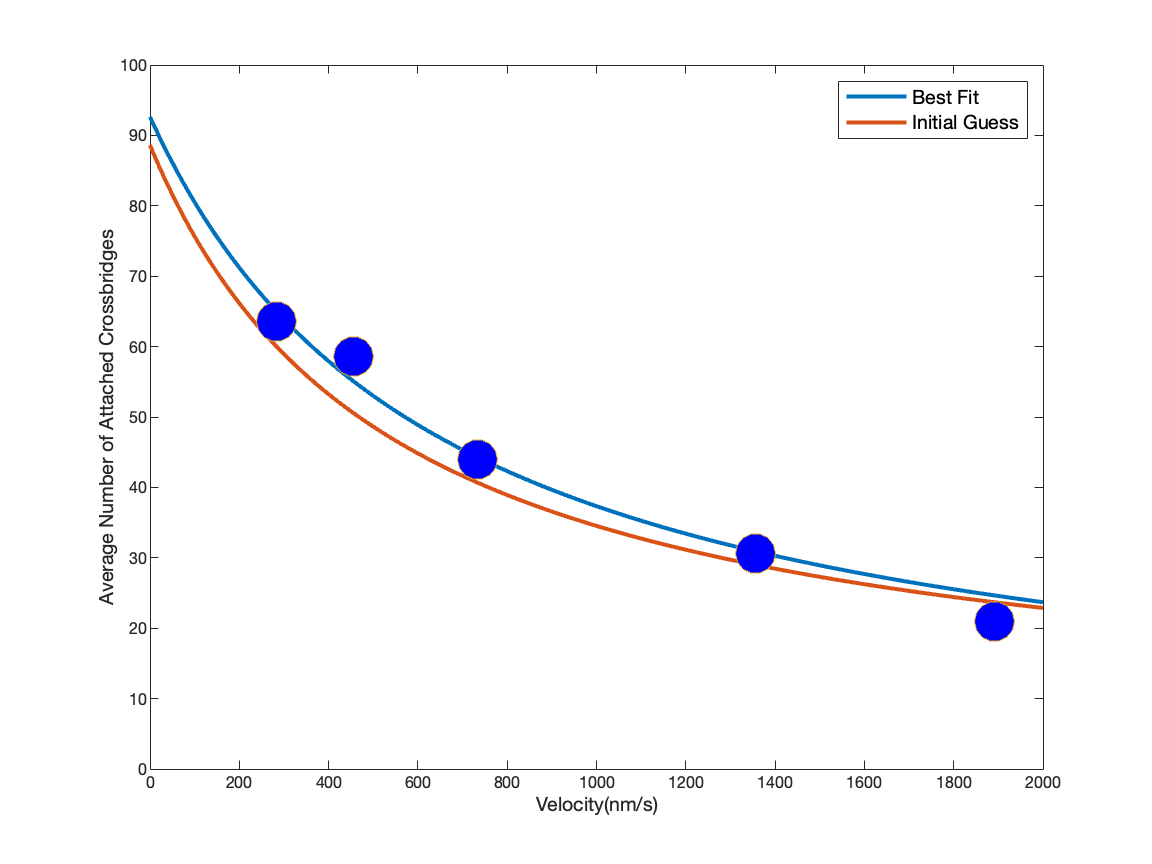}
   \includegraphics[scale=0.19]{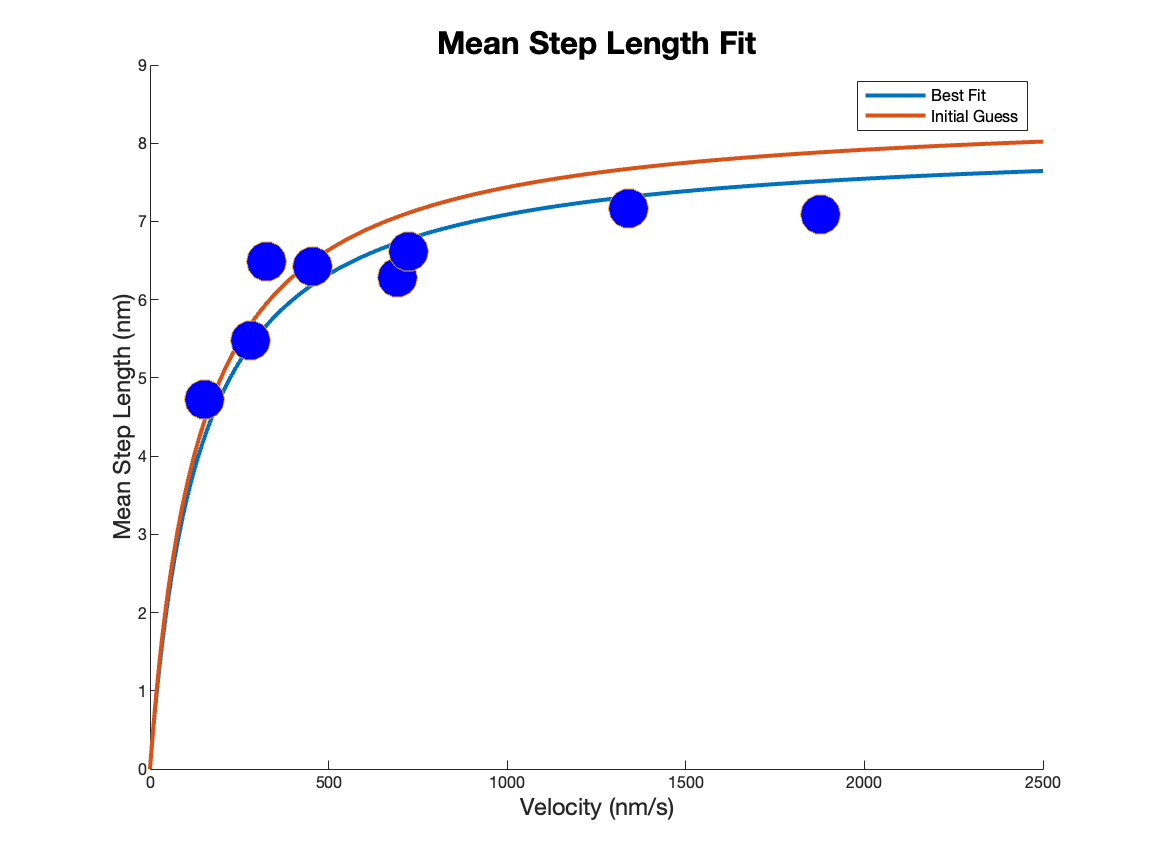}
  \caption{Left panel: Fitting \eqref{eq:force_fit_eq} and \eqref{eq:fit_force_num} to the data in Figure 1c of \cite{PIAZZESI2007784}. Middle panel: Fitting \eqref{eq:U(v)_finite} to the data in Figure 4a of \cite{PIAZZESI2007784}. Right panel: Fitting \eqref{eq:mean_step_length} to Figure 4c of \cite{PIAZZESI2007784}. In all panels, data points in \cite{PIAZZESI2007784} are plotted as blue dots. The red curves appeared in all panels (i.e. the right-sides of the arrows appear in the first row of Table~\ref{table:parameters}) are the fit of our initial guess whereas the blue curves are our best fits to the exact steady-state theory of $ k = 3.3$pN/nm (i.e. the second row of Table~\ref{table:parameters}). }
  \label{fig:fit_1}
\end{figure}

The fitting results are reported in graphical form in Figure~\ref{fig:fit_1}.  They show that the fit
 is already good when the model is simplified by taking the limit $k \rightarrow \infty$, and that the fit is even better when the measured value of
 $k=3.3$ pN/nm is used.

\section{Series Elasticity}

Up to now we have been considering a muscle that is shortening
at a constant velocity,  and this has been sufficient to determine
the parameters of our model.  In experiments, a constant velocity
of shortening is achieved by applying a constant load force to
the contracting muscle.  In this scenario, it makes no difference
whether the load force is applied directly to the contractile
apparatus, or through a (linear or nonlinear) series elastic
element, since the length of any such element will be constant
when the load force is constant.  In more general experiments,
however, if there is a series elastic element between the
contractile machinery and the load, this will influence the
dynamics of contraction.

The need for a series elastic element is apparent from the
quick-release experiments reported in~\cite{podolsky1960kinetics}.  In those experiments, an
activated muscle is first held at constant length until its force
becomes steady, and then the muscle is suddenly allowed to begin
shortening against a load that is smaller than the steady isometric
force.  When the force on the muscle suddenly decreases, there is a
corresponding downward jump in length.  The change in length divided
by the change in force is a measure of the compliance (reciprocal of
stiffness) of the whole system. Part of this compliance can be
attributed to the crossbridges themselves, but it turns out that the
compliance of the crossbridges is too small to account for the size of
the jump that is seen.  This suggests that there is additional
compliance in series with that of the crossbridges (compliances in
series add).  Although we cannot say where within the muscle this
additional compliance resides, a reasonable guess would be in the z-line
apparatus that connects one sarcomere to the next.  In that case, half
of the compliance of each z-line can be considered to be the
compliance in series with each half-sarcomere.

The basic equations that are needed to take series elasticity
into account are as follows.  First, the total length of the
half sarcomere is now
\begin{equation}
    L = l + \lse,
\end{equation}
where $l$ is the length of the contractile apparatus and $\lse$ is the
length of the series elastic element.  The force on the two parts are
the same, and we denote this force by $\mathcal{P}$.  Note that $\mathcal{P} = N_c P$ is equal to the sum of the forces generated by
 the attached crossbridges in a half sarcomere.  Here $N_c$ is the
 number of cycling crossbridges in a half sarcomere, and $P$ is
 the expected value of the force generated by any one cycling
 crossbridge (regardless of attachment status). 
When the muscle is connected to a constant load, as in the
isotonic phase of a quick-release experiment,  $\mathcal{P}$ is also equal to the
given load, and this is a constraint that the crossbridge forces must
satisfy.

We assume that the series elastic element is a linear spring, and this
gives the equation
\begin{equation}
    \mathcal{P} = \kse \lse
\end{equation}
We do not include a rest length, since it would be constant anyway,
and we are only concerned with changes in length.

As already mentioned, $\mathcal{P}$ is constant during the isotonic phase of
a quick-release experiment, and it then follows that $\lse$ is constant
as well.  For this reason, the series elastic element has no effect
on the dynamics on the isotonic phase.

During the isometric phase, however, the series elastic element
does change the behavior of the muscle.  This is because the
isometric condition is now that $l+\lse$ is constant. Then
\begin{equation}
    d\lse/dt = - dl/dt = v(t)
\end{equation}

and therefore
\begin{equation}
       d\mathcal{P}/dt = \kse v(t),
\end{equation}
and of course v(t) is no longer constrained to be zero, so it needs to
be determined.  Also, every time that a crossbridge detaches, there is
a sudden decrease in $\mathcal{P}$, and this causes a corresponding decrease in
$\lse$, leading to an increase in $l$ so that $l+\lse$ remains constant.
Thus, because of the series elastic element, the isometric muscle
experiences crossbridge shortening between detachment events, and
crossbridge lengthening (of the crossbridges that remain attached) at
each crossbridge detachment event.

Also, the transition from the isometric phase to the isotonic phase of
the quick-release experiment is influenced by the series elastic
element, which changes length as a result of the sudden decrease in
force.

These effects of the series elastic element on the isometric and transition phases of a quick release experiment are described in more detail in the following section.

\section{Simulation Methodology}

In this section, we describe methodology for computer simulation of a
 quick-release experiment~\cite{podolsky1960kinetics}, in which the muscle is held at a
 constant length for a period of time, during which it develops a
 certain level of force, which is known as the isometric force, and
 then the muscle is released from the constant-length constraint and
 is allowed to shorten against a force that is smaller than the
 isometric force.  The constant-length part of the experiment is
 called the isometric phase, and the part in which the muscle is
 shortening against a smaller constant force is called the isotonic
 phase.  Between these two phases is an instantaneous decrease in
 length that we call the transition phase.

Our simulation methodology is stochastic.  It follows the individual
 crossbridges in a half-sarcomere as they undergo the crossbridge
 cycle of attachment, force development, and detachment.  For a
 previous example of this kind of simulation applied to a different
 crossbridge model, see~\cite{duke1999molecular}.  We use event-driven simulation, in
 which the events are those of crossbridge attachment and detachment.
 For any un-attached crossbridge, attachment occurs with a fixed
 probability per unit time, $\alpha$.  For any attached crossbridge, the
 probablility per unit time of detachment depends in our model on the
 crossbridge force through the function $\beta(p)$, and this is
 time-dependent, since the crossbridge force p depends on time, and in
 a different way for every attached crossbridge, since the different
 crossbridges were attached at different times.

The simulation jumps from one event to the next.  After each event, a
 random attachment time is assigned to each un-attached crossbridge,
 and a random detachment time to each attached crossbridge.  Out of
 all of these $\textit{possible}$ next events, the one that is scheduled to occur
 earliest is the one that $\textit{actually}$ occurs.  The assignment of the
 random attachment and detachment times is done by drawing each such
 time from the appropriate probability distribution.  For attachement
 this is simple (see Appendix A) because the probability per unit time
 of attachment is constant, but for detachment it is complicated and
 requires the evaluation of $\beta(p(t))$ for each attached crossbridge
 separately during every inter-event interval.  How this is done is
 described below.  The details are different and are therefore
 described separately for the isometric phase and for the isotonic
 phase of the simulation.  Once the functions $\beta(p(t))$ have been
 evaluated for each of the attached crossbridges, the method of using
 each of those functions to choose a random detachment time for that
 crossbridge is described in Appendix A.

An advantage of this kind of stochastic simulation in comparison to
 the formulation and numerical solution of differential equations that
 describe the crossbridge population is that event-driven simulation
 has no timestep parameter $\Delta t$.  Instead, the timescale of the
 simulation is set by the events themselves.  Indeed, event-driven
 simulation has no numerical error; it generates an exact realization
 of the stochastic process defined by the mathematical model
 \footnote{Strictly speaking, the statement that event-driven
simulation is exact ignores three sources of error.  One of these is
that computer arithmentic involves roundoff error.  But the relative
errors in roundoff are about $10^{-16}$, which is $\textit{much}$ smaller
than typical discretization error that arises in the numerical
solution of differential equations.  Another source of error is that
random-number generators are not truly random, but a lot of work has
gone into making them do a good job of generating numbers with the
same statistical properties as if they were random.  Finally, in our
particular application, Newton's method needs to be used in the choice
of the random detachment times for the attached crossbridges, and this
involves numerical error, but the quadratic convergence behavior of
Newton's method (doubling the number of correct digits at each step as
the solution is approached) makes this source of error negligible as
well.}.  Another feature of event-driven simulation that may be
 considered an advantage or a disadvantage is that it automatically
 generates the intrinsic noise of the process.  This is an advantage
 if that noise is of interest.  If not, then there is the disadvantage
 that averaging of multiple simulations may be required to remove the
 noise.  In our application, however, the crossbridge population of a
 half-sarcomere is large enough that the noise of a single simulation
 is not overwhelming, and by doing just a few repetitions we can
 easily get a sense of what the mean behavior of the model is, and
 also how much variability.  This will become apparent in the Results
 section of the paper.

A quick-release experiment consists of three phases:

\begin{enumerate}
    \item \textbf{Isometric Phase}: During the isometric phase, the presence of series elasticity (see
 previous section) complicates the dynamics.  Instead of $l =$ constant,
 we have $l+l_{se} =$ constant.  Here we derive the equations that
 govern the behavior of the muscle in the presence of this constraint.
 First, we consider what happens during time intervals in which there
 are no attachment or detachment events, and then we evaluate the
 instantaneous changes that occur during any attachment or detachment
 event. Attachment is simple because the
 newly attached crossbridge has zero force, so there is no jump in
 force or length when a crossbridge attaches. There is, howeverm a jump in v because the number of attached
 crossbridges has changed, see equation~\eqref{eq:76}, below.  When a crossbridge detaches,
 however, there is a sudden decrease in force because the force that
 was applied by the detaching crossbridge is lost, and because this is
 only partially compensated by stretch of the crossbridges that remain
 attached.  Thus, crossbridge detachment during the isometric phase is
 accompanied by jumps in $l$ and in $l_{se}$, with $l+l_{se}$ of course
 remaining constant. These jumps are evaluated in the following, along
 with the resulting jump in force.

 The equations that hold during any time interval in which there is
 no crossbridge attachment or detachment are as follows.  Since
 $l + l_{se}$ is constant, and since $v = -dl/dt$, 
\begin{equation}
\label{eq:lse_1}
	\frac{d \lse}{dt} = - \frac{d l}{dt} = v(t)
\end{equation}
\begin{equation}
\label{eq:elasticity_force}
	\frac{d \mathcal{P}}{d t} = \kse v(t)
\end{equation}

Moreover, it follows from \eqref{eq:ODE_force} that 
\begin{equation}
\label{eq:long_eq}
	\frac{d \mathcal{P}}{d t} = \sum_{i\in A} k \left(\vmax (1- \frac{p_i}{p_\infty}) - v \right) = N_A k \left(\vmax (1- \frac{\mathcal{P}}{p_\infty N_A}) - v \right)
\end{equation}
where $A$ is the set of indices of attached crossbridges, and where $N_A$ is the number of attached crossberidges. 
Then, by combining \eqref{eq:elasticity_force} and \eqref{eq:long_eq}, 
\begin{equation}
\label{eq:long_eq_2}
\kse v	= N_A k \left(\vmax (1- \frac{\mathcal{P}}{p_\infty N_A}) - v \right)
\end{equation}
which is equivalent to 
\begin{equation}
\label{eq:ODE_total_force}
	\frac{d}{dt} (1 - \frac{\mathcal{P}}{N_A p_\infty}) = - \frac{\kse k N_A \vmax}{(\kse + k N_A) N_A p_\infty} (1 - \frac{\mathcal{P}}{N_A p_\infty})
\end{equation}
As we are considering here the crossbridge dynamics within a time interval between two consecutive crossbridge attachment/detachment events, let $t_0$ denote the time of the most recent crossbridge
 attachment of detachment event, and let $t_1$ denote the
 time of the next such event. Solving the ODE \eqref{eq:ODE_total_force} explicitly gives 
\begin{equation}
\label{eq:P_solution}
	(1 - \frac{\mathcal{P} (t) }{N_A p_\infty}) =  (1 - \frac{\mathcal{P} (t_0^+)}{N_A p_\infty}) \exp(-\lambda (t - t_0))
\end{equation}
where $t \in (t_0,t_1)$ and $\lambda = \frac{\kse k N_A \vmax}{(\kse + k N_A) N_A p_\infty} = \frac{\kse (k N_A)}{\kse + (k N_A)} \frac{\vmax}{N_A p_\infty}$. Note that $\frac{\kse (k N_A)}{\kse + (k N_A)}$ is the stiffness of two springs in series, one with stiffness $\kse$ and the other with stiffness $k N_A$, which is the aggregated stiffness of $N_A$ attached crossbridges in parallel. 

Notice from \eqref{eq:long_eq_2} that 
\begin{equation}
\label{eq:76}
	v(t) = \vmax \frac{k N_A}{\kse + k N_A} (1 - \frac{\mathcal{P} (t) }{N_A p_\infty})
\end{equation}
Hence, 
\begin{equation}
\label{eq:elastic_v}
	v(t) = \vmax \frac{k N_A}{\kse + k N_A}(1 - \frac{\mathcal{P} (t_0^+)}{N_A p_\infty}) \exp(-\lambda (t - t_0)) 
\end{equation}
Evaluating $v(t)$ at $t = t_0^+$ gives
\begin{equation}
\label{eq:v_t0_plus}
	v(t_0^+) = \frac{N_A p_\infty - \mathcal{P} (t_0^+)}{\kse} \lambda
\end{equation}
\begin{equation}
	v(t) = v(t_0^+) \exp(-\lambda (t - t_0))
\end{equation}

With these equations, it is convenient to rewrite \eqref{eq:ODE_force} as 
\begin{equation}
\label{eq:force_lambda}
	\frac{d}{dt} (1 - \frac{p}{p_\infty}) = - (\frac{k \vmax}{p_\infty}) \left[(1 - \frac{p}{p_\infty}) -  \frac{v(t_0^+)}{\vmax} \exp(-\lambda (t - t_0))\right]
\end{equation}
In this equation, p is the force of any one of the crossbridges that happen to be attached during the time interval $(t_0,t_1)$.
Let $\mu = \frac{k \vmax}{p_\infty}$ and it follows immediately that $\lambda = \frac{\kse (k N_A)}{\kse + (k N_A)} \mu < \mu$. In terms of $\mu $, \eqref{eq:force_lambda} becomes 
\begin{equation}
\label{eq:force_mu}
	\frac{d}{dt} \left(1 - \frac{p}{p_\infty} \exp(-\lambda \mu (t - t_0))\right) =\mu \frac{v(t_0^+)}{\vmax}\exp\left(-\lambda (\mu-\lambda) (t - t_0)\right)
\end{equation}

Integrating \eqref{eq:force_mu} over $(t_0,t)$ yields 
\begin{equation}
	1 - \frac{p}{p_\infty} = (1 - \frac{p(t_0^+)}{p_\infty}) e^{-\mu (t -t_0)} + \frac{N_A p_\infty - \mathcal{P} (t_0^+)}{\kse \vmax} (\frac{\mu \lambda}{\mu - \lambda}) \left(e^{-\lambda (t -t_0)} - e^{(-\mu (t -t_0)}\right)
\end{equation}

The probability per unit time of crossbridge detachment is given by \eqref{eq:beta_final} and we let
\begin{equation}
\label{eq:bar_beta}
\begin{aligned}
    \Bar{\beta} (\tau) & = \beta (p (t_0 + \tau)) \\ & = \frac{\alpha}{4} \left[1 + 20 \left ((1 - \frac{p(t_0^+)}{p_\infty}) e^{-\mu \tau} +  \frac{N_A p_\infty - \mathcal{P} (t_0^+)}{\kse \vmax} (\frac{\mu \lambda}{\mu - \lambda}) \left(e^{-\lambda \tau} - e^{-\mu \tau}\right)\right)\right] 
\end{aligned}
\end{equation}

Note that this is a different function for each attached crossbridge because of the parameter $p(t_0^+)$. The function $\Bar{\beta} (\tau)$ will be used to determine the random detachment time of each attached crossbridge during the isometric state with the event-driven simulation method described in Appendix A.  

Next we consider the consequences of series elasticity when a crossbridge
 attaches or detaches during the isometric phase, so that there is no
 change in muscle length. If $t_0$ is the time of a crossbridge attachment event, then the newly attached crossbrdige has zero force. It follows that, for the newly attached crossbridge, $p(t_0^+) = 0$. For all crossbridges that were already attached before the event at $t_0$, $p(t_0^+) = p(t_0^-)$, and therefore $\mathcal{P} (t_0^+) = \mathcal{P} (t_0^-)$. Moreover, we have the relations: $N_A(t_0^+) = N_A(t_0^-) + 1$, $l(t_0^+) = l(t_0^-)$, $\lse(t_0^+) = \lse(t_0^-)$.  Thus, during an attachment event, all forces and lengths are continuous, but the discrete variable $N_A$ increases by $1$. Note that this implies that $v$ is not continuous, because of \eqref{eq:76}. 

If $t_0$ is instead the time of a detachment event, the situation is more complicated. Let $*$ denote the index of the crossbridge that detaches, so that $p_\ast (t_0^-)$ is the crossbrdige force that is lost as a result of detachment. As a result of this loss of crossbridge force, an unknown jump in length, denoted by $\Delta l$ occurs. Since we are in the isometric phase, it follows from \eqref{eq:lse_1} that the change of length in $\lse$ is equal to minus the change of length in $l$:
\begin{equation}
	\Delta \lse = - \Delta l
\end{equation}

Then, we conclude from \eqref{eq:elasticity_force} that 
\begin{equation}
\label{eq:Delta_P}
	\Delta \mathcal{P} = - \kse \Delta l
\end{equation}
 
 On the other hand, by considering the crossbridge stiffness $k$ and the number of bridges that remain attached $N_A (t_0^+)$, we also have 
 \begin{equation}
 \label{eq:Delta_P_2}
 	\Delta \mathcal{P} = - p_\ast (t_0^-) + k N_A (t_0^+)\Delta l
 \end{equation}
 
 Then, we can solve for $\Delta l$ and $\Delta \mathcal{P}$ with \eqref{eq:Delta_P} -\eqref{eq:Delta_P_2}:
 \begin{equation}
 	\Delta l = \frac{p_\ast (t_0^-)}{\kse + k N_A (t_0^+)}
 \end{equation}
 
 \begin{equation}
 	\Delta \mathcal{P} = - \frac{\kse p_\ast (t_0^-)} {\kse + k N_A (t_0^+)}
 \end{equation}
 
 Also, for each individual crossbridge that was attached just prior to $t = t_0$ and remains attached immediately after $t = t_0$, 
 \begin{equation}
 	\Delta p = k \Delta l = \frac{k p_\ast (t_0^-)}{\kse + k N_A (t_0^+)}
 \end{equation}
 
 Following either type of event during the isometric phase, the functions $\Bar{\beta} (\tau)$ that are given by \eqref{eq:bar_beta} become different from what they were before. This is not only because $N_A$ has changed, but also because the functions $\Bar{\beta} (\tau)$ depend on the parameters $p(t_0^+)$ and $\mathcal{P}(t_0^+)$. Whether or not these parameters jump at $t = t_0$, they will typically have different values from the values that they had immediately following the previous attachment or detachment event. 
 
    \item \textbf{Transition Phase}: 
After the muscle has arrived at a statistical steady state in which
 the total crossbridge force is fluctuating around a steady value denoted by $F_0$, we release the muscle from its constant-length constraint, and we allow it to shorten agains a given load $F_* < F_0$.  During the jump in force from $F_0$ to $F_*$,
 the muscle instantaneously shortens.  There are two contributions to this shortening, one from the series elastic element, and the other from the crossbridges themselves.  The shortening that occurs, together with the change in force of each attached crossbridge,
 are evaluated as follows:
    
    We use the same notation $t_0$ for the time of the event of transition. The isotonic phase is characterized by a prescribed force $F\ast$. Neither attachment nor detachment occurs during the transition, so $N_A (t_0^+) = N_A (t_0) = N_A (t_0^-)$. We will use the notation $N_A (t_0)$ to denote all these quantities as they are always identical. The change of total force during the transition is given by 
 \begin{equation}
 	\Delta \mathcal{P} = F_\ast - \mathcal{P} (t_0^-) < 0
 \end{equation}
 
 This results in 
 \begin{equation}
 	\Delta \lse = \frac{\Delta \mathcal{P}}{\kse}
 \end{equation}
 \begin{equation}
 	\Delta \l = \frac{\Delta \mathcal{P}}{k N_A (t_0)}
 \end{equation}
 
 Therefore, the jump in length of the whole system during the transition is 
 \begin{equation}
 	\Delta \lse + \Delta l = \Delta \mathcal{P} (\frac{1}{\kse} + \frac{1}{k N_A (t_0)})
 \end{equation}
 
 Meanwhile, because of the change $\Delta l$, each attached crossbridge has a change in force given by 
 \begin{equation}
 	\Delta p = k \Delta l = \frac{\Delta \mathcal{P}}{N_A (t_0)}
 \end{equation}

    \item \textbf{Isotonic Phase}:  
    During the isotonic phase, the load force has the constant value $F_*$,
 and this immediately implies that the series elastic element has
 constant length.  Thus the changes in muscle length occur exactly
 as if the series elastic element were not present.  Also, the
 total crossbridge force has to be equal to F* at all times.
 We have to consider how this condition is maintained during
 the time intervals between attachment/detachment events, and
 also how it is maintained during those events themselves.

Qualitatively, the muscle has to shoten during the time intervals
 between detachment/attachment events.  This is to compensate for the
 internal crossbridge dynamics which would be increasing the
 crossbridge force if the muscle length were not changing.  During a
 crossbridge detachment event, there is an abrupt \textit{increase} in
 muscle length, so that the increase in force of the crosbridges that
 remain attached will just compensste for the force that was lost
 during detahcment.  During an attachment event, there is no change in
 muscle length, since the force of the newly attached crossbridge is
 zero.  This qualitative discussion is made quantitative in the
 following.

    While the muscle crossbridges detach/attach during the simulation with rates $\beta(p)$ and $\alpha$, the attached crossbridges move towards $x > 0$ to compensate for the change of total force and to keep the total force fixed at $F_\ast$. Each crossbridge has a probability per unit time of detaching $\beta (p)$, which is different for different crossbridges. At the same time, unattached crossbridges attach at $x = 0$ with rate $\alpha$. There is again an equilibrium state that the crossbridges will arrive at. In this equilibrium, all attached crossbridges move with a speed that fluctuates around a constant value and the number of attached crossbridges also fluctuates around a constant value. Whenever a new crossbridge attaches, there is no jump in the muscle length because the attached crossbridge has an initial force of zero. However, when a detachment event occurs, there is an upward jump in
 the muscle length, thereby stretching all the crossbridges that are
 still attached by just the right amount to maintain a constant force
 $F_\ast$. So what happens overall is intervals of shortening interrupted by
 abrupt upward jumps in muscle length.

    	After the transition phase, the evolution of crossbridge forces is described by equation \eqref{eq:ODE_force}. Summing equation \eqref{eq:ODE_force} over all attached crossbridges, we have that 
    	\begin{equation}
\label{eq:sum_force_isotonic}
    \sum_{i\in A}  \frac{dp_i}{dt} = \sum_{i\in A} k \left(v_\text{max} (1-\frac{p_i}{p_\infty}) - v\right)
\end{equation}

Notice that we keep the force $F = \sum_i A_i p_i = F_\ast$, which is constant during the isotonic phase. This means that the left-hand side of equation \eqref{eq:sum_force_isotonic} is zero. Hence, we can rewrite equation \eqref{eq:sum_force_isotonic} as\begin{equation}
\label{eq:vel_between_isotonic}
    v = v_\text{max} (1 - \frac{F_\ast}{p_\infty N_A})
\end{equation}
Remarkably, this shows that $v$ is constant during the isotonic phase during any time interval in which there is no crossbridge attachement or detachment occurs. However, if there is a crossbridge attached or detached, then the change of the resulting $N_A$ will lead to the change of $v$. There is a difference still between a detachment event and an attachment event. During an attachment event, $v$ jumps but there is no jump in length. During an detachment event, $v$ jumps and there is a jump in length.

Since $v$ is constant, and given by \eqref{eq:vel_between_isotonic}, we can obtain an equation
 for $p_i$ by sustituting \eqref{eq:vel_between_isotonic} into \eqref{eq:ODE_force}, with $p$ replaced by $p_i$ in \eqref{eq:ODE_force}:

\begin{equation}
\label{eq:force_between_isotonic}
    \frac{dp_i}{dt} = \frac{k v_\text{max}}{p_\infty} (\frac{F_\ast}{N_A} - p_i) 
\end{equation}
This is easily solved for $p_i(t)$. Let $t_E$ be the time when a crossbridge attaches or detaches. Then, until the next event occurs, for $t > t_E$, the solution of \eqref{eq:force_between_isotonic} is 
\begin{equation}
    p_i (t) = \frac{F_\ast}{N_A} + (p_i(t_E^{+}) - \frac{F_\ast}{N_A}) \exp(-\frac{k v_\text{max}}{p_\infty} (t -t_E))
\end{equation}
where $t_E^{+}$ is the moment right after a crossbridge attaches or detaches. 

The upward jump in muscle length that occurs during the isotonic
 phase whenever a crossbridge detaches can be evaluated as
 follows.  Let $i^\ast$ be the index of the crossbridge that is detaching.
 Then, for any crossbridge that remains attached
\begin{equation}
\label{eq:pos_jump_isotonic}
     x(t_E^{+}) - x(t_E^{-}) = \frac{p_{i^{\ast}}(t_E^{-})} {k N_A(t_E^{+}) }
\end{equation}
 and this is also the change in length of the half sarcomere.



\end{enumerate}

\begin{figure}[h!]
   \includegraphics[scale=0.3]{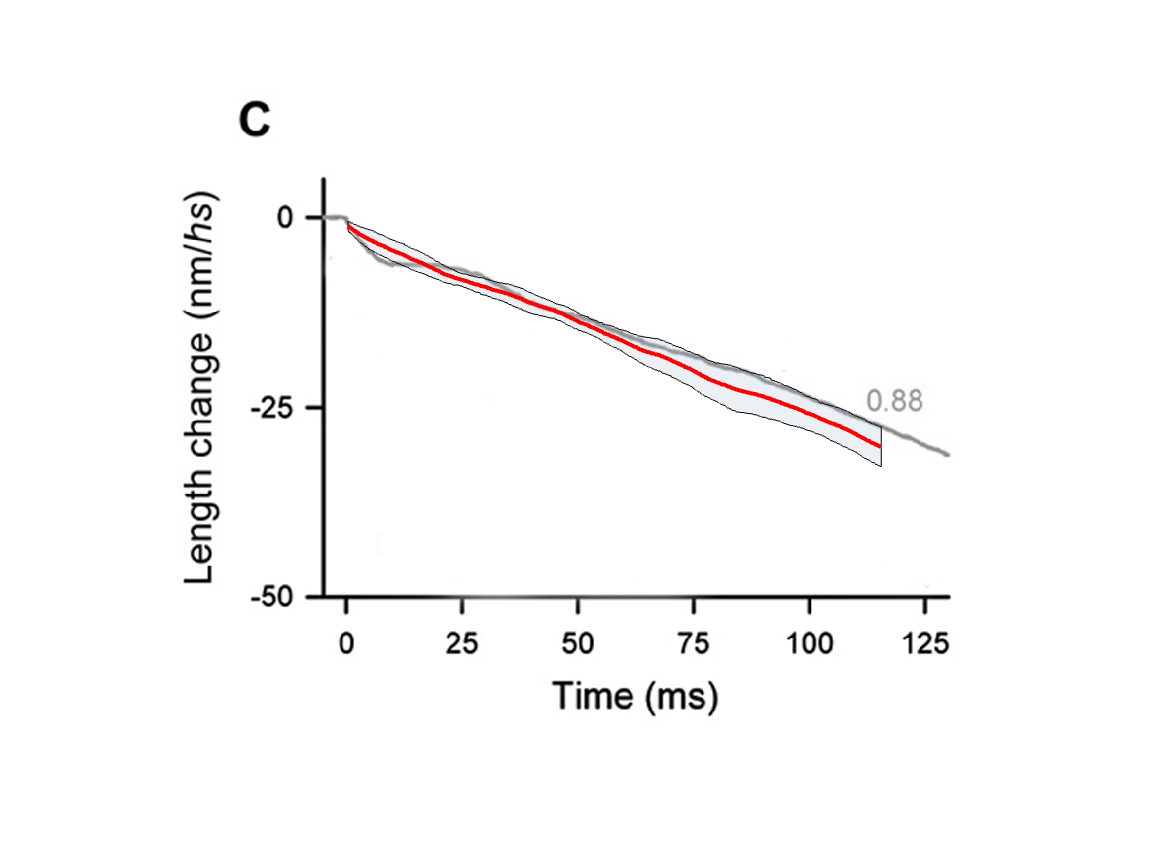}
   \includegraphics[scale=0.3]{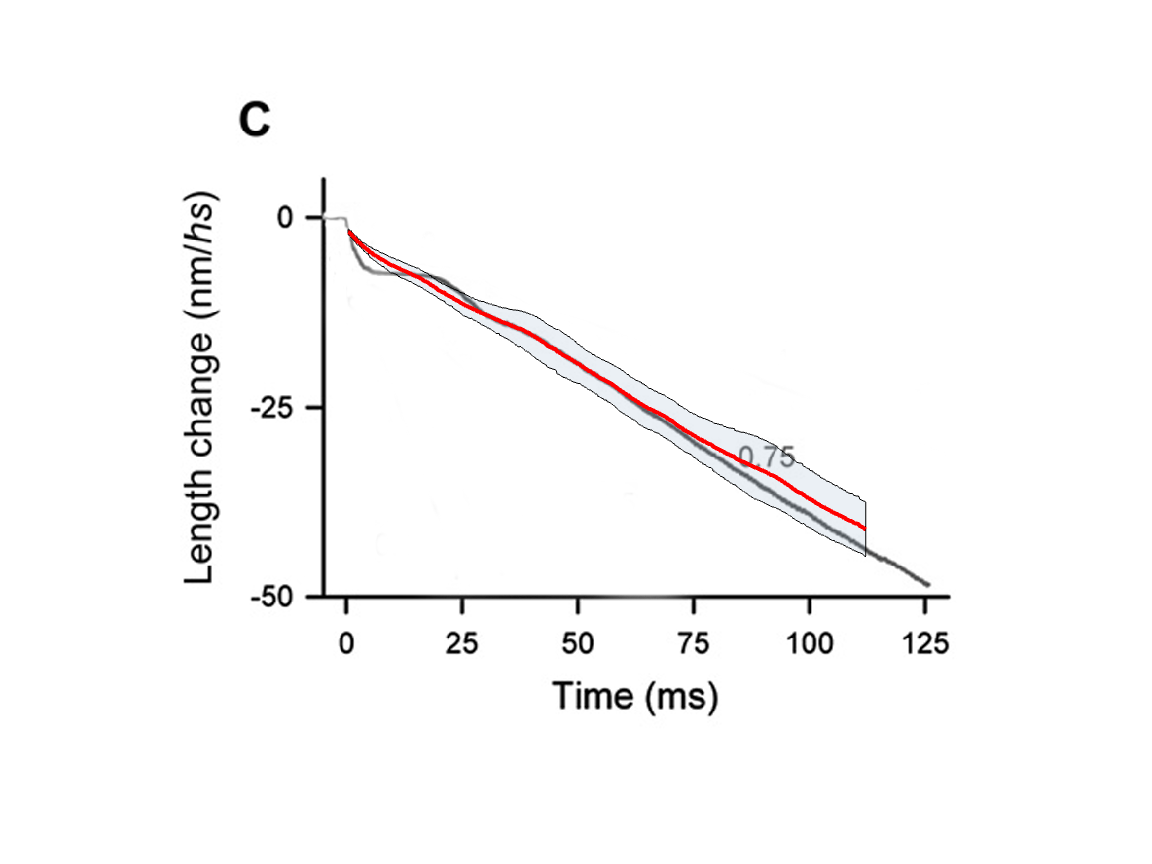}
      \includegraphics[scale=0.3]{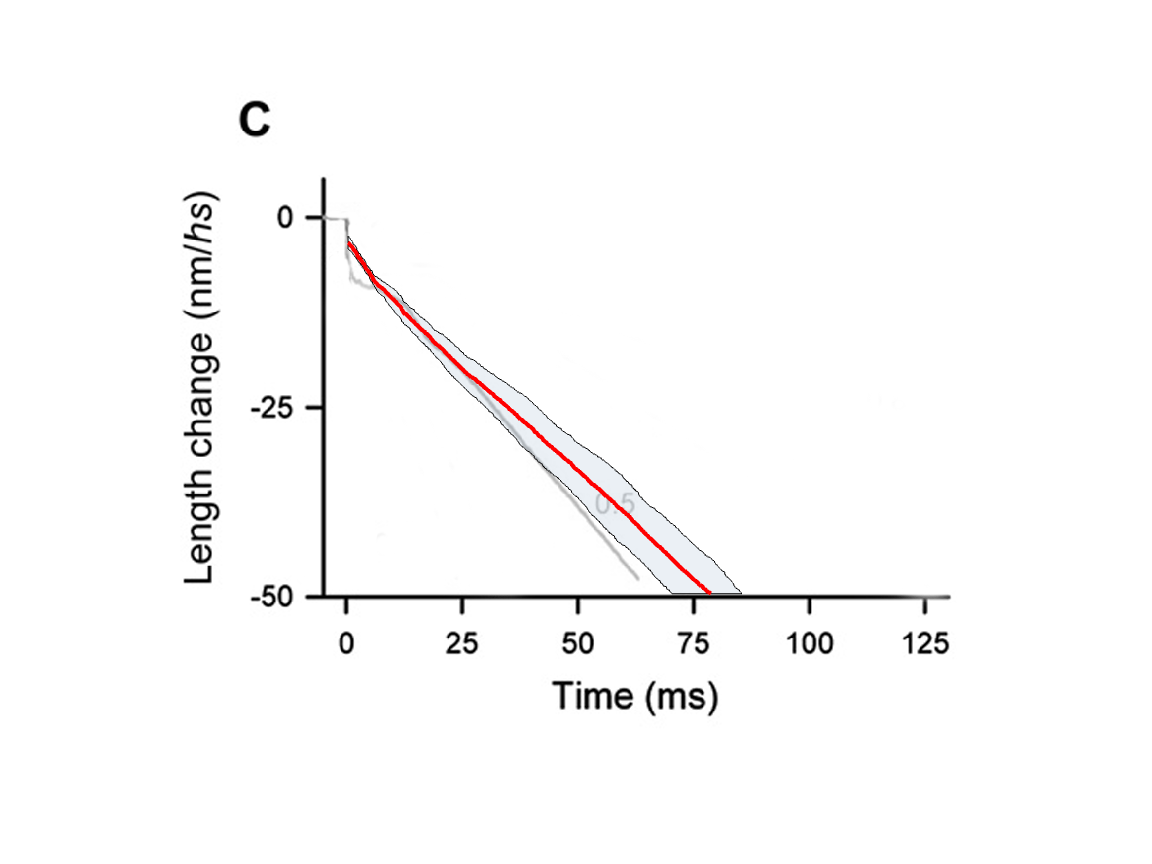}
   \includegraphics[scale=0.3]{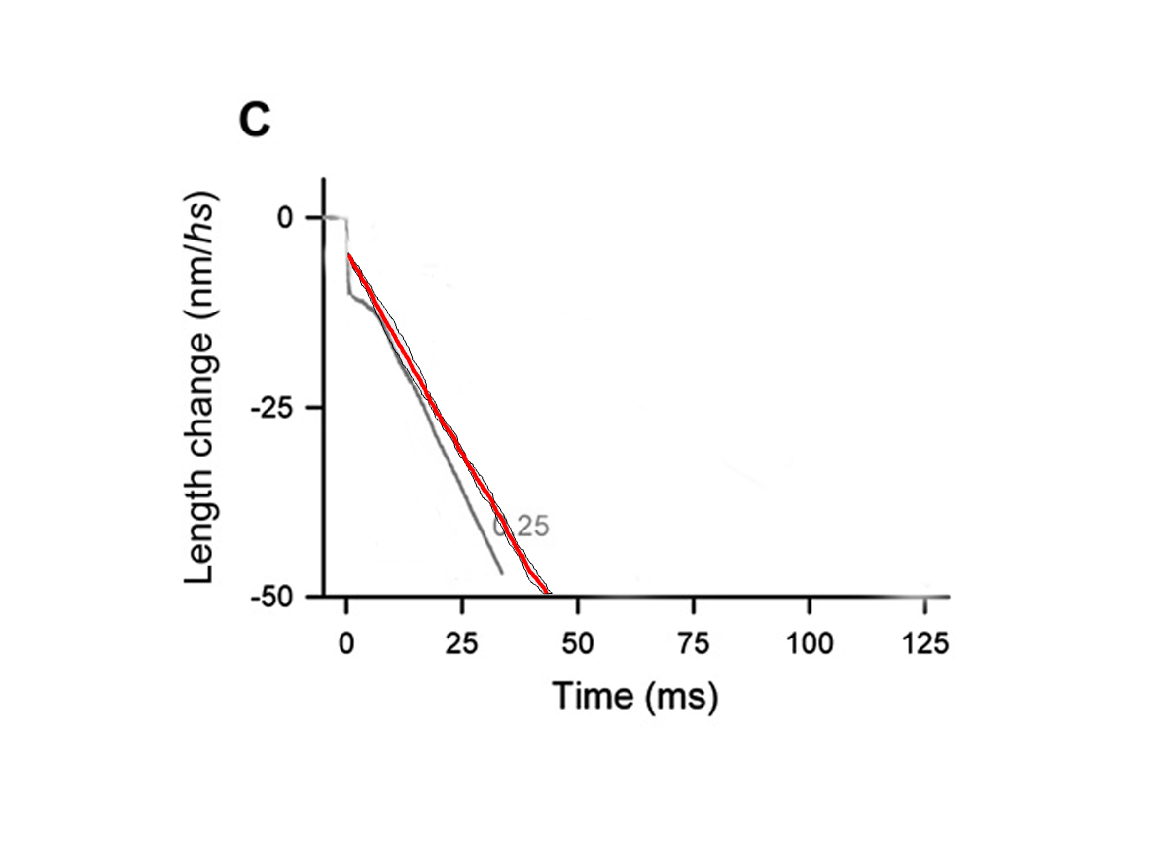}
      \includegraphics[scale=0.3]{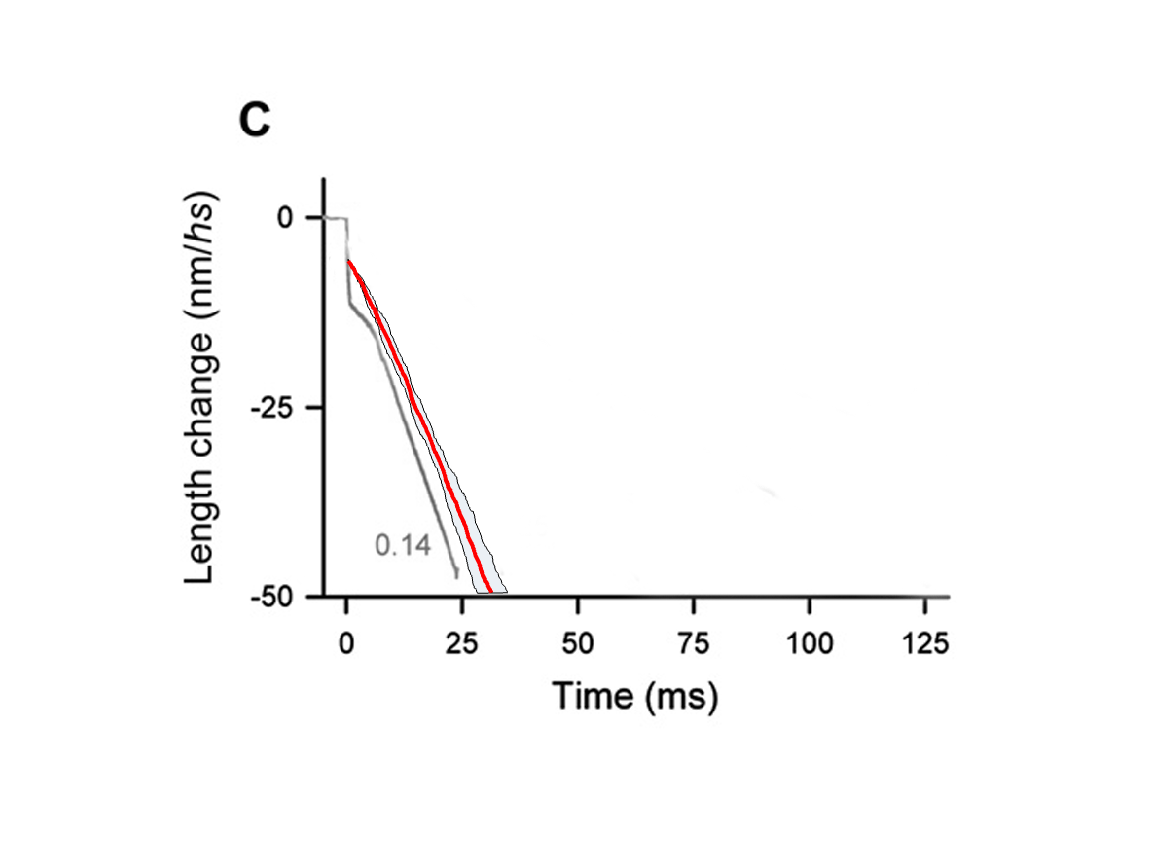}
  \caption{Quick release simulations compared to experimental results.
The plots show length vs time, with the zero of time at the instant of
release, and with the zero of length being the pre-release length, so
that all lengths are negative.  The lengths shown are in "nm/hs",
which denotes nanometers per half sarcomere.  The figures were
reproduced from~\cite{PIAZZESI2007784}, and then the simulated data were plotted on top
of the experimental data to the same scale.  The original experimental
data are shown as gray lines.  A red line shows the average of 10
simulation results in each case, and the shaded regions centered on
the red lines show $\pm$ one standard deviation around the average of
the simulated data.  All parameters of the model except for
$k_\text{se} = 100 \text{pN/nm}$ were already determined from
steady-state considerations (see Section 5) before undertaking these
quick-release simulations.  The number in each panel is the
post-release load on the muscle expressed as a fraction of the
(pre-release) isometric force.} 
  \label{fig:simulation results}
\end{figure}

\begin{figure}[t!]
  \centering      
         \includegraphics[scale=0.27]{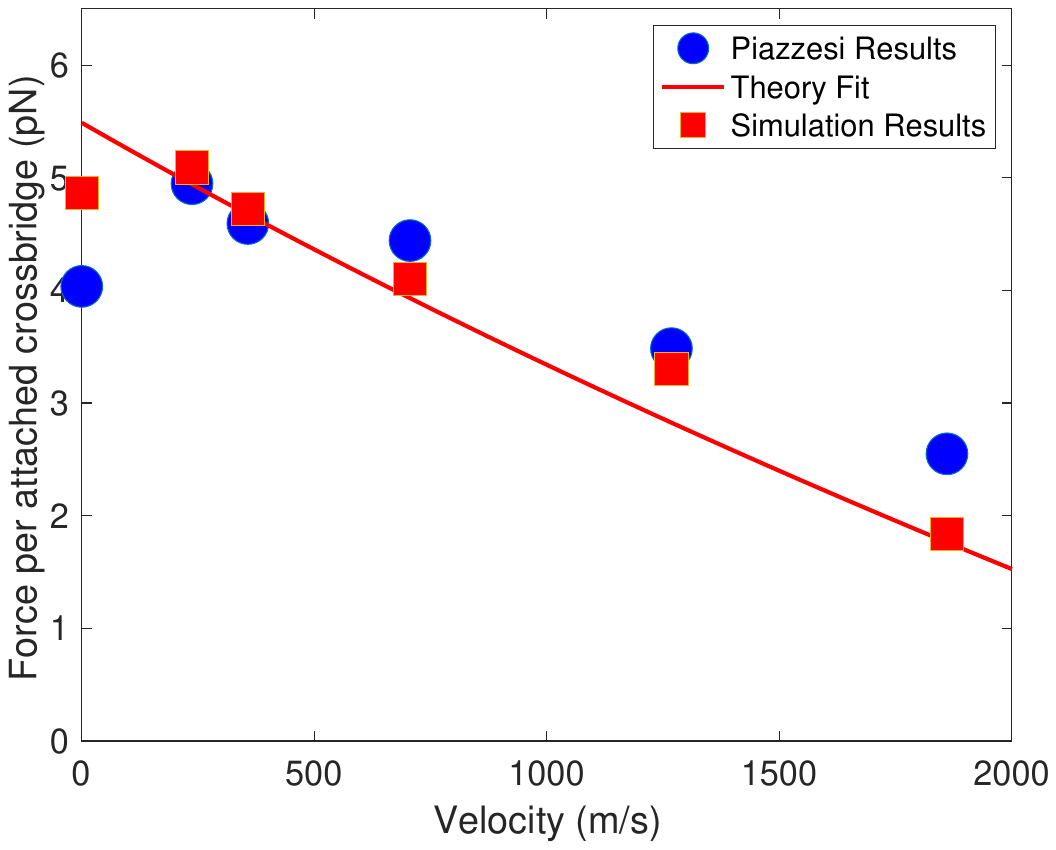}
   \includegraphics[scale=0.27]{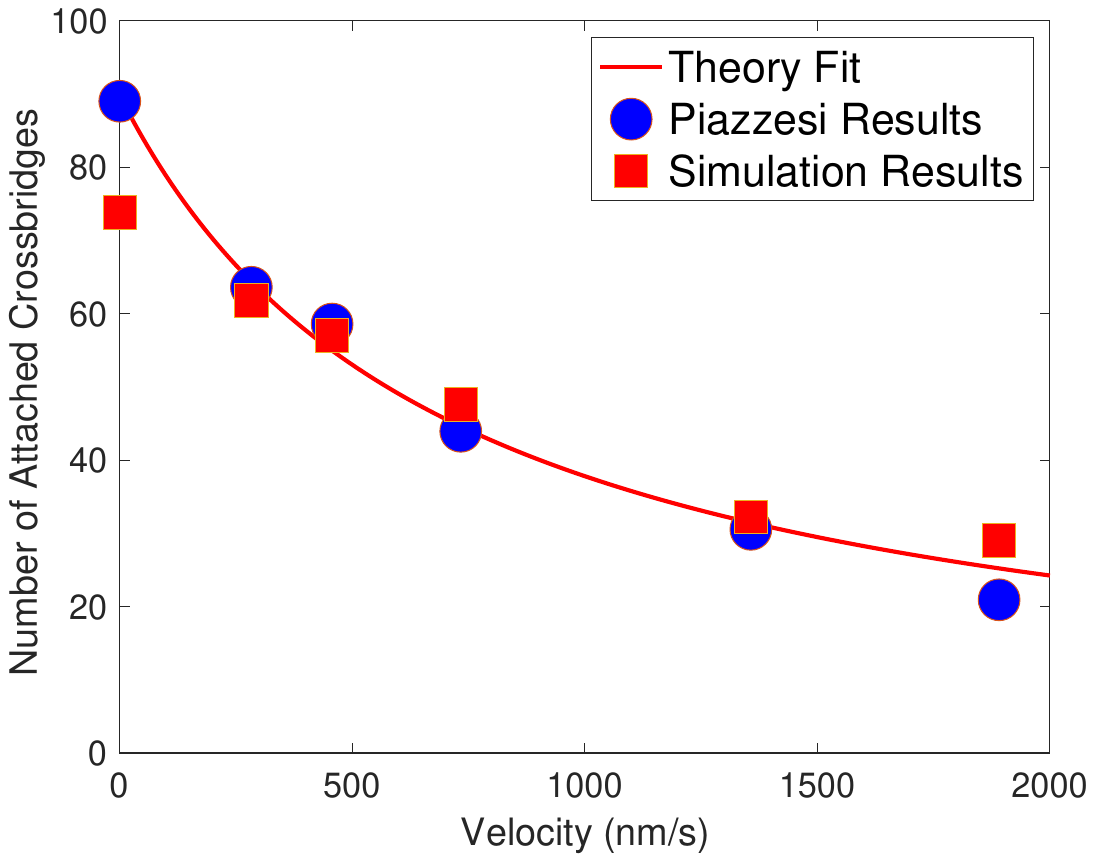}
  \caption{Steady-state simulation results (red filled squares)
compared to steady-state experimental results (blue filled circles)
and also to the exact steady-state theory of Section 4 (red lines).
The left panel shows the force per attached crossbridge, and the right
panel shows the number of attached crossbridges, in both cases as a
function of the steady velocity of shortening.} 
  \label{fig:simulation results_ex}
\end{figure}

\section{Simulation Results}

With the methodology and numerical algorithms detailed in the previous
section, we simulate the quick release experiments reported by
Piazzesi [10].  The protocol of a quick release experiment is described
at the beginning of the previous section.

All of the parameters except $k_\text{se}$, which is the series elasticity
of a half sarcomere, have already been determined, see Section 5.
In principle, $k_\text{se}$ can be determined by fitting to the size of
the sudden change in length that occurs when the muscle is allowed to
begin shorten against a given load, so the force decreases instantaneously
from the isometric force to that of the load.
\footnote{In applying this method of determining $k_\text{se}$, it is
important to recall that the overall stiffness of the half-sarcomere,
i.e., the change in force divided by the change in length during
the jump from the isometric state to the isotonic state,
is given by $k_\text{se} N_A k / (k_\text{se} + N_A k)$, where $N_A$
is the number of attached crossbridges and k is the stiffness of
each attached crossbridge.}
In practice, however, it is hard to discern in the experimental record
where the true jump in length ends and a subsequent period of rapid
shortening begins.  So instead of doing any fitting we just make the
arbitrary choice of $k_\text{se} = 100$pN/nm.  This choice makes the
series elasticity of a half-sarcomere be equal to the aggregate
stiffness of 30 attached crossbridges.

The results of our simulations are shown in Figure~\ref{fig:simulation results}, superimposed on
the experimental records, reproduced from~\cite{PIAZZESI2007784}.  Each result (red line
in the figure) is the average of 10 simulations, and the shaded area
around the red line has a width of one standard deviation on either
side of the average.  The gray line in each case is the correspondng
experimental result.  The five cases shown differ in the load against
which the muscle is allowed to shorten starting at time t=0, after
being held at constant length during $t<0$.  These isotonic loads are
the following fractions of the isometric force: $0.88, 0.75, 0.5, 0.25,
0.14$.  In all cases, the experimental line lies within or near one
standard deviation from the average of the simulation results.
Agreement is better for the three largest loads than for the two
smallest loads.  The most obvious differences between experimental and
simulation results occur during the rapid transient that
immediately follows $t = 0$.  In comparison to the experimental
results, the simulations show what looks like a smaller instantaneous
decrease in length at $t=0$, and less oscillation during the subsequent
approach to a constant velocity of shortening.  Note, however, that
the overall downward displacement of the shortening trajectory
associated with the reduction in force at $t = 0$ comes out about right.
That is, if the straight-line part of the plot of length vs time is
extrapolated back to time zero, the intercept on the length axis is in
approximately the same place for the experimental and for the
corresponing simulation result in each case.  From this point of view,
the experimental data appear to be following a damped oscillatory
approach to our simulation data.  The mechanism underlying these
damped oscillations is unexplained by our model, but one may
speculate that inertial effects are involved.

\section{Summary and Conclusions}
The motivation for this paper has been to create a crossbridge theory
that resolves the apparent contradiction that a crossbridge acts
like a linear spring, and yet the force generated by an attached
crossbridge is fairly constant during most of the time interval in
which it is attached \cite{PIAZZESI2007784}, despite changes in overall extension of the
crossbridge caused by relative motion of the thick and thin
filaments.  What this suggests is that the crossbridge is able
to "take up the slack" and thereby maintain a relatively constant strain
during sarcomere shortening.

We have created a crossbridge theory of this kind.  In our theory, the
crossbridge spring has constant stiffness, but its rest length is an
internal dynamical variable, with its own (linear) relationship
between force and velocity.  Crossbridges attach to the thin filament
in a state of zero strain, and then strain, and therefore force, is
developed at a finite rate by the internal dynamics of the
crossbridge.  According to our theory, the maximum velocity of
shortening (which occurs at zero load) is determined by the maximum
rate at which the rest length of the crossbridge can change.  This is
in sharp contrast to the original Huxley theory~\cite{huxley1957}, in which
crossbridges attach in a strained configuration, and in which the
shortening of unloaded muscle involves a balance of forces between a
part of the crossbridge population that is trying to shorten the
muscle, and another part that is trying to lengthen it.  Our model has
the more appealing property that in normal muscle function all attached
crossbridges are pulling and none are pushing, regardless of the
velocity of shortening.

An assumption of our model is that the probability per unit time
of crossbridge detachment is a function of the crossbridge force.
To determine this function, we consider a limiting case of the
model, in which the crossbridge stiffness becomes infinite.
In this limiting case we observe that the model can be put
in perfect agreement with the original discoveries of A.V. Hill~\cite{AVHILL}
concerning the force-velocity curve and heat of shortening
of contracting muscle, if and only if we choose a detachment rate
that is a linearly \textit{decreasing} function of the
crossbridge force.  Although perhaps unexpected, this "catch-bond"
behavior may be useful from a functional point of view, since
it maintains crossbridge attachment for a longer time when the
crossbridge is making a larger contribution to the force generated
by the muscle.

With the form of our crossbridge model completely determined,
we turn to the task of parameter fitting.  For this purpose,
we return to the case of finite crossbridge stiffness
and derive exact steady-state results that can be compared
to experimental data for the purpose of setting the paramerers.
With the parameters thus determined, it turns out that
the crossbridge stiffness is indeed large in an appropirate
dimensionkess sense, and this justifies our use of the limiting
case of infinite crossbridge stiffness to determine the rate
of detachment, as described above.

To go beyond steady-state applications of the model, we introduce an
event-driven stochastic simulation method for the crossbridge dynamics
problem, and we apply this method to the simnulation of a
quick-release experiment, with comparison to the experimental data of
\cite{PIAZZESI2007784}.  This requires the introduction of a series elastic element,
since the stiffness of the crossbridges themselves is too large (i.e.,
their compliance is too small) to account for the size of the downward
jump in length that ocurs when the muscle is suddenly allowed to begin
shortening against a smaller than isometric load.  Except for the
series elasticity, we fix all of the parameters of the model at the
values that were determined from steady-state considerations, and even
for the series elasticity, we choose one particular value somewhat
arbitrarily, and do not do any parameter fitting.  Nevertheless, our
simulations of the quick release experiment show good agreement with
the data of~\cite{PIAZZESI2007784}, not only with respect to the slope of the straight
line that approximates the plot of length vs time once the velocity
has become steady, but also with respect to the intercept of that line
extrapolated back to t=0.  Despite this overall agreement, there are
differences of detail in the experimental and simulated results.  The
experimental results show a damped oscillatory appoach to the steady
state, and these oscillation are almost absent from the simulated
results.  This may point to a limitation of our model, but the
possibility that the oscillations are being produced by inertia in the
experimental case should also be considered.

The introduction of series elasticity has a peculiar consequence that
it produces a fundamental distinction between the isometric state and
an isotonic state in which the load happens to be such that the mean
shortening velocity is zero.  In any isotonic state, the length of the
series elastic element is constant, and this implies that the series
elasticity has no influence on the dynamics.  In the isometric state,
however, fluctuations in force produce fluctuations in length of the
series elastic element, and this allows for (indeed, requires)
fluctuating relative motion of the thick and thin filaments.
In our model, this subtle effect provides a partial explanation
for the experimental anomaly that the force per attached crossbridge
in isometric muscle is quite different from what would be expected
by extrapolating the isotonic force per attached crossbridge to
a load that produces zero velocity, see Figure~\ref{fig:simulation results_ex}, left panel,
in which both the experimental result and the corresponding
result of our simulation are plotted.  Regardless of the
validity of the above explanation, the experimental anomaly
is striking and very much worthy of further investigation.

In summary, we hope that this paper provides a fresh perspective on
crossbridge dynamics, and that it will encourage both theoretical and
experimental studies in which the internal state of the crossbridge
plays a central role.

\section{Appendix A}
In this Appendix, we explain how we obtain samples of random times $T_\text{a}$ and $T_\text{d}$ from equation \eqref{eq:rate_alpha} and \eqref{eq:rate_beta}. 

\subsection{Sampling from a random time with a constant probability per unit time}

By solving the ODE \eqref{eq:rate_alpha} directly, we obtain the following relation:

\begin{equation}
	\text{Pr}(T_a > \tau) = \exp(- \int_{0}^{t} \alpha d\tau') = \exp(- \alpha \tau)
\end{equation}
and the CDF of the resulting distribution of $T_a$ is 
\begin{equation}
	F(t) = 1 - \text{Pr}(T_a > \tau) = 1 - \exp(- \alpha \tau)
\end{equation}
where we notice that $T_a$ follows an exponential distribution of coefficient $\alpha$ and the sampling of such a distribution is simple: we take a uniform random variable $U$ between $[0,1]$, and our sample $T_a$ is 
\begin{equation}
	T_a = - \log(U) /\alpha
\end{equation}

\subsection{Sampling from a random time with a variable probability per unit time}

This section aims to sample from the probability distribution of $T_d$ defined by equation \eqref{eq:rate_beta}. Let $\bar\beta (\tau) = \beta (p(\tau))$. Note that $\beta (p(\tau))$ has an explicit expression so we can find $\bar\beta (\tau)$ in each case needed. Similar to the previous section, we integrate the equation \eqref{eq:rate_beta} and we get
\begin{equation}
	\text{Pr} (T_d > \tau) = \exp(- \int_{0}^\tau \bar\beta(\tau') d\tau' ) 
\end{equation}

We draw a random variable $U$ from the uniform distribution between $[0,1]$ and we notice that 
\begin{equation}
	\text{Pr} (U < u) = u
\end{equation}
for any $u \in [0,1]$. Then, to find $T_d$, we solve the nonlinear equation 
\begin{equation}
\label{eq:variable_coef_newton}
	\exp( - \int_{0}^\tau \bar\beta(\tau') d\tau' ) =  U
\end{equation}

We note that the solution $t = T_d$ to this equation is unique since the left-hand side is a strictly increasing function from $[0,1]$, given the assumption that $\int_{0}^1 \bar\beta(\tau') d\tau' = \infty$ and the right-hand side is bounded. The assumption can be easily verified through checking \eqref{eq:beta_final}, which gives the lower bound that $\beta > \frac{\alpha}{4}$. 

Now, we need to consider the numerical solution of \eqref{eq:variable_coef_newton} and it can be rewritten as 
\begin{equation}
	\label{eq:rewrite_newton}
	 \int_{0}^\tau \bar\beta(\tau') d\tau' + \log U = 0
\end{equation}
The solution of \eqref{eq:rewrite_newton} by Newton's method can be obtained via the iteration 
\begin{equation}
	T_{n+1} = T_n - \frac{\int_{0}^\tau \bar\beta(\tau') d\tau' + \log U}{\bar\beta(\tau)}
\end{equation}
and the resulting solution from the Newton's iteration can be chosen as a sample of $T_d$. 

\section{Appendix B}

In 1938, A.V. Hill\cite{AVHILL} summarized the results of his experimental investigation on contracting skeletal muscle in the following way. 

First, when the muscle is shortening against a constant load force, denoted $P$, the velocity of shortening, denoted by $v$, is given by

\begin{equation}
\label{eq:Hill_app_1}
	v = b \frac{P_0 - P}{a + P}
\end{equation}

in which $a,b$ and $P_0$ are constants.

Next, the rate at which the muscle generates heat is given by 
\begin{equation}
\label{eq:Hill_app_2}
	\dot{H} = M_0 + av
\end{equation}
The constant $M_0$ is called the maintenance heat, and the term $av$ is called the heat of shortening. 

Note the remarkable fact that the same parameter $a$ appears in both of the above equations. Its value is given by 
\begin{equation}
	a = \frac{P_0}{4}
\end{equation}

Another remarkable relationship between equations \eqref{eq:Hill_app_1} and \eqref{eq:Hill_app_2} is that the velocity is at which $M_0 = av$ is the same as the velocity that maximizes the power delivered by the muscle to the load. In general, the rate at which the muscle does work on the load is given by 
\begin{equation}
	\dot{W} = vp = b \frac{(P_0 - P)P}{a + P}
\end{equation}
and this is maximized when 
\begin{equation}
	0 = (a + P) (P_0 - 2P) - (P_0 - P) p = a P_0 - 2aP - P^2
\end{equation}
or 
\begin{equation}
	(\frac{P}{P_0})^2 + 2 \frac{a}{P_0} \frac{P}{P_0} - \frac{a}{P_0} = 0
\end{equation}

The positive solution of this quadratic equation is 
\begin{equation}
\label{eq:hill_positive}
	\frac{P}{P_0} = - (\frac{a}{P_0}) + \sqrt{(\frac{a}{P_0})^2 + (\frac{a}{P_0})}
\end{equation}

The velocity corresponding to this optimal load is obtained by substituting \eqref{eq:hill_positive} into \eqref{eq:Hill_app_1} with the result
\begin{equation}
\label{eq:hill_vel1}
    v = \frac{1 + \frac{a}{P_0} - \sqrt{(\frac{a}{P_0})^2 + (\frac{a}{P_0})}}{\sqrt{(\frac{a}{P_0})^2 + (\frac{a}{P_0})}}
\end{equation}

This can be put in a nicer form by introducing
\begin{equation}
    \vmax = \frac{b P_0}{a}, 
\end{equation}
which is the velocity of shortening by zero load. Then, \eqref{eq:hill_vel1} becomes
\begin{equation}
\label{eq:hill_v_vmax}
    \frac{v}{\vmax} = - \frac{a}{P_0} + \sqrt{(\frac{a}{P_0})^2 + (\frac{a}{P_0})}
\end{equation}

Note that the right-hand sides of \eqref{eq:hill_positive} and \eqref{eq:hill_v_vmax} are the same! This apparent coincidence is a consequence of the fact that, when equation \eqref{eq:Hill_app_1} is rewritten in terms of $\vmax$ instead of $b$, it can be put in the form 
\begin{equation}
\label{eq:hill_one}
    (\frac{v}{\vmax})(\frac{P}{P_0})(\frac{P_0}{a}) + (\frac{v}{\vmax}) + (\frac{P}{P_0}) = 1
\end{equation}
which is symmetrical in $(\frac{P}{P_0})$ and $(\frac{v}{\vmax})$. In our case $\frac{a}{P_0} = \frac{1}{4}$, so the optimal $v$, which is the value of $v$ at which $M_0 = av$ is given by 
$\frac{v}{\vmax} = \frac{1}{4} (\sqrt{5} - 1)$. Note that $\sqrt{5}$ is very close to $\frac{9}{4}$. It will therefore simplify all of our formulae without making any practical difference if we choose 
\begin{equation}
    M_0 = a \vmax \frac{1}{4} (\frac{9}{4} - 1) = a \vmax \frac{5}{16}
\end{equation}
Then \eqref{eq:Hill_app_2} becomes 
\begin{equation}
    \dot{H} = a (\frac{5}{16} \vmax + v) = \frac{P_0 \vmax}{4} (\frac{5}{16}  + \frac{v}{\vmax})
\end{equation}
Note that $\dot{H}$ is fully determined, and we can evaluate the rate at which the muscle is consuming chemical energy as 
\begin{equation}
    \dot{E} = \dot{W} + \dot{H}
\end{equation}
To express this as a function of $v$, we need $P$ as a function of $v$. From \eqref{eq:hill_one}, 
\begin{equation}
\label{eq:hill_p_p0}
    \frac{P}{P_0} = \frac{1 - \frac{v}{\vmax}}{1 + (\frac{v}{\vmax})(\frac{P_0}{a})} = \frac{1 - \frac{v}{\vmax}}{1 + 4(\frac{v}{\vmax})}
\end{equation}
Then, since $\dot{W} = vp$, 
\begin{equation}
\label{eq:hill_dot_e}
    \dot{E} = P_0 \vmax \left( \frac{v}{\vmax}\frac{1 - \frac{v}{\vmax}}{1 + 4(\frac{v}{\vmax})} + \frac{1}{4}(\frac{5}{16} + \frac{v}{\vmax})\right) = \frac{5}{64} P_0 \vmax \frac{1 + 20 \frac{v}{\vmax}}{ 1+ 4 \frac{v}{\vmax}}
\end{equation}
Since each crossbridge cycle involves the hydrolysis of the molecule of ATP, it must be the case that the rate of crossbridge cyclying is proportional to $\dot{E}$, and this explains equation \eqref{eq:Hill2} of the main text. Also, since $P_0$ is the value of $P$ when $v = 0$, equation \eqref{eq:hill_p_p0} of this appendix is the same as equation \eqref{eq:Hill1} in the main text. 

The above description of the mechanics and energetics of skeletal muscle contraction is certainly not exact. It has been refined by Hill himself\cite{hill1964effect}, and by many other investigators \cite{Alcazar2019OnTS}. Within the framework of the crossbrdige model of the present paper, these results are exact only in the limit of infinite crossbridge stiffness. Since the crossbridge stiffness turns out to be large in an dimensionless sense, this provides a mechanistic explanation of Hill's empirical observations. 

\bibliographystyle{siam}
\bibliography{reference}

\begin{thebibliography}{10}

\bibitem{alcazar2019shape}
{\sc J.~Alcazar, R.~Csapo, I.~Ara, and L.~M. Alegre}, {\em On the shape of the force-velocity relationship in skeletal muscles: The linear, the hyperbolic, and the double-hyperbolic}, Frontiers in physiology, 10 (2019), p.~438208.

\bibitem{Alcazar2019OnTS}
{\sc J.~Alcazar, R.~Csapo, I.~Ara, and L.~M. Alegre}, {\em On the shape of the force-velocity relationship in skeletal muscles: The linear, the hyperbolic, and the double-hyperbolic}, Frontiers in Physiology, 10 (2019).

\bibitem{duke1999molecular}
{\sc T.~Duke}, {\em Molecular model of muscle contraction}, Proceedings of the National Academy of Sciences, 96 (1999), pp.~2770--2775.

\bibitem{EISENBERG1980195}
{\sc E.~Eisenberg, T.~Hill, and Y.~Chen}, {\em Cross-bridge model of muscle contraction. quantitative analysis}, Biophysical Journal, 29 (1980), pp.~195--227.

\bibitem{AVHILL}
{\sc A.~V. Hill}, {\em The heat of shortening and the dynamic constants of muscle}, Proceedings of the Royal Society of London. Series B - Biological Sciences, 126 (1938), pp.~136--195.

\bibitem{hill1964effect}
\leavevmode\vrule height 2pt depth -1.6pt width 23pt, {\em The effect of load on the heat of shortening of muscle}, Proceedings of the Royal Society of London. Series B. Biological Sciences, 159 (1964), pp.~297--318.

\bibitem{hill1974theoretical}
{\sc T.~L. Hill}, {\em Theoretical formalism for the sliding filament model of contraction of striated muscle part i}, Progress in biophysics and molecular biology, 28 (1974), pp.~267--340.

\bibitem{huxley1957}
{\sc A.~Huxley}, {\em 6 - muscle structure and theories of contraction}, Progress in Biophysics and Biophysical Chemistry, 7 (1957), pp.~255--318.

\bibitem{lacker1986mathematical}
{\sc H.~M. Lacker and C.~Peskin}, {\em A mathematical method for the unique determination of cross-bridge properties from steady-state mechanical and energetic experiments on macroscopic muscle}, in Lectures on mathematics in the life sciences, AMS, 1986, pp.~121--153.

\bibitem{PIAZZESI2007784}
{\sc G.~Piazzesi, M.~Reconditi, M.~Linari, L.~Lucii, P.~Bianco, E.~Brunello, V.~Decostre, A.~Stewart, D.~B. Gore, T.~C. Irving, M.~Irving, and V.~Lombardi}, {\em Skeletal muscle performance determined by modulation of number of myosin motors rather than motor force or stroke size}, Cell, 131 (2007), pp.~784--795.

\bibitem{podolsky1960kinetics}
{\sc R.~Podolsky}, {\em Kinetics of muscular contraction: the approach to the steady state}, Nature, 188 (1960), pp.~666--668.

\bibitem{walcott2012mechanical}
{\sc S.~Walcott, D.~M. Warshaw, and E.~P. Debold}, {\em Mechanical coupling between myosin molecules causes differences between ensemble and single-molecule measurements}, Biophysical journal, 103 (2012), pp.~501--510.

\end{thebibliography}

\end{document}